\UseRawInputEncoding
\documentclass[USenglish, twocolumn]{article}

\usepackage{graphicx}
\usepackage{adjustbox}
\usepackage{amssymb}
\usepackage{hyperref}
\usepackage{caption}
\usepackage{subcaption}
\usepackage{titling}

\title{\textbf{Design and Fabrication of Robust Hybrid Photonic Crystal Cavities}}

\author{Alex Abulnaga$^{1,\nmid}$, Sean Karg$^{1,\nmid}$, Sounak Mukherjee$^{1}$, Adbhut Gupta$^{1}$, Kirk W. Baldwin$^{1}$, Loren N. Pfeiffer$^{1}$, Nathalie P. de Leon$^{*,1}$}

\date{\vspace{-0.5cm} \today}

\renewcommand{\maketitle}{
    \begin{center}
        \vspace{-2cm} 
        {\huge \thetitle \par}
        \vspace{0.5cm}
        
        {\normalsize \theauthor \par}
        \vspace{0.2cm}
        
        {\itshape
        \textit{$^1$Department of Electrical and Computer Engineering, Princeton University, Princeton, USA} \\ 
        \textit{$^\nmid$ These authors contributed equally to this work} \\
        $^*$\textit{Corresponding author: npdeleon@princeton.edu}
        \par}
        
        \vspace{1cm}
        {\normalsize \thedate \par}
        \vspace{-0.25cm}
    \end{center}
}

\begin{document}

\twocolumn[
  \begin{@twocolumnfalse}
    \maketitle
    \begin{abstract}
    Heterogeneously integrated hybrid photonic crystal cavities enable strong light-matter interactions with solid-state, optically addressable quantum memories. A key challenge to realizing high quality factor (Q) hybrid photonic crystals is the reduced index contrast on the substrate compared to suspended devices in air. This challenge is particularly acute for color centers in diamond because of diamond's high refractive index, which leads to increased scattering loss into the substrate. Here we develop a design methodology for hybrid photonic crystals utilizing a detailed understanding of substrate-mediated loss, which incorporates sensitivity to fabrication errors as a critical parameter. Using this methodology we design robust, high-Q, GaAs-on-diamond photonic crystal cavities, and by optimizing our fabrication procedure we experimentally realize cavities with Q approaching 30,000 at a resonance wavelength of 955~nm.
    \end{abstract}
    \vspace{0.75cm}
\end{@twocolumnfalse}
]
    
\section{Introduction} 
Color centers and other atom-like defects in the solid state are a promising platform for long distance quantum networks because they can have long spin coherence times and efficient spin-photon interfaces, and they can be incorporated into scalable photonic devices \cite{atature_material_2018, zhang_material_2020}. Some of the most sophisticated quantum network demonstrations to date are based on color centers in diamond such as the nitrogen-vacancy (NV) \cite{pompili_realization_2021, stolk_telecom-band_2022, bersin_telecom_2024, stolk_metropolitan-scale_2024} and silicon-vacancy (SiV) centers \cite{bhaskar_experimental_2020, knaut_entanglement_2024}, as well as rare earth ions in various host materials \cite{dibos_atomic_2018, ruskuc_nuclear_2022, ourari_indistinguishable_2023, uysal_spin-photon_2024, ruskuc_scalable_2024}. These qubits can be incorporated into nanophotonic cavities to enhance light-matter interactions and achieve improved spin-photon entanglement rates \cite{reiserer_colloquium_2022}. One approach to cavity integration is to mill \cite{zhong_nanophotonic_2015,  kindem_control_2020} or etch \cite{burek_free-standing_2012, sipahigil_integrated_2016, guo_tunable_2021, ding_high-q_2024} monolithic cavities out of the host material, but this can introduce surface defects and subsurface damage that can degrade the spin and optical properties of the qubit \cite{chu_coherent_2014, riedel_deterministic_2017, siyushev_optical_2017, bhaskar_quantum_2017, guo_direct-bonded_2023, ruf_resonant_2021}. An alternative approach is to fabricate the photonic device using another material that is well-suited for photonics fabrication, and then couple the qubit to the evanescent field by placing the cavity on the substrate containing the qubit, allowing the qubit to reside in a more pristine environment. This strategy has been recently deployed for Er ions in various host materials  \cite{raha_optical_2020, chen_parallel_2020, dibos_purcell_2022, huang_stark_2023, ourari_indistinguishable_2023, yu_frequency_2023, horvath_strong_2023}, Yb ions in YVO$_4$ \cite{wu_near-infrared_2023}, and SiV centers in diamond \cite{chakravarthi_hybrid_2023}. The key challenge for hybrid photonic crystal cavities is scattering into the high index substrate, which leads to lower Q as compared to devices in air. This is particularly challenging for material systems with a low index contrast and operation wavelengths in the visible range. While hybrid devices with Q up to 190,000 have been demonstrated for silicon (n=3.5) on CaWO$_4$ (n=1.9) in the telecom band \cite{ourari_indistinguishable_2023}, for lower index contrast devices such as GaP-on-diamond \cite{chakravarthi_hybrid_2023}, the demonstrated Q is limited to less than 10,000 in the visible wavelength range.

The typical methodology for designing 1D photonic crystals is to maximize the expected Purcell factor by designing a structure with a mode volume (V) close to the minimum possible value, and then maximizing Q. This can be accomplished by maximizing the photonic band gap and lengthening the tapered region that defines the cavity mode, and then iteratively optimizing the cavity parameters to achieve the highest Q/V. However, despite hybrid device designs that can achieve Q exceeding 1 million in simulation \cite{huang_hybrid_2021, zhou_photonic_2022, wu_near-infrared_2023}, these high Qs have not been experimentally realized in fabricated devices.

Here we demonstrate that robustness to fabrication errors is a critical parameter in designing photonic crystal cavities that is independent of the designed Q. We perform random sampling of cavity designs under realistic fabrication errors and observe a significant scatter in the simulated error-sensitivity across designs. Armed with this design methodology, we then show that previously demonstrated hybrid photonic crystal cavities are likely limited by fabrication errors, not by sidewall roughness, absorption losses, or other scattering mechanisms. We focus on GaAs-on-diamond \cite{huang_hybrid_2021} as a model system that is suitable for coupling to neutral SiV centers in diamond \cite{rose_observation_2018}. In order to fabricate devices with the correct target parameters, we optimize the lithography, etching, and undercut chemistry to realize GaAs-on-diamond hybrid photonic crystals with Q approaching 30,000 at a resonance wavelength of 955~nm and exceeding 43,000 at 1520~nm. We hypothesize that these demonstrations have now achieved material-absorption-limited Q.

\section{Hybrid photonic crystal design}
The key figure of merit in designing an optical cavity is the Purcell enhancement of the emitter spontaneous emission rate, given by $P = 4g(\vec{r})^2/\kappa \Gamma_0$, where $\Gamma_0$ is the native spontaneous emission rate of the optical transition of interest, $g(\vec{r})$ is the single-photon Rabi frequency, and $\kappa$ is the cavity decay rate. The Rabi frequency is determined by the overlap between the cavity electric field, $\vec{E}(\vec{r})$, and the emitter dipole-moment, $\vec{\mu}$, as $g(\vec{r}) = \vec{\mu}\cdot\vec{E}(\vec{r})/\hbar$. The cavity decay rate of a resonance at frequency $\omega$ is defined as $\kappa = \omega/Q$ where $Q$ is the quality factor of the resonance. To maximize the Purcell factor, we seek cavity designs that achieve resonances with high Q and concentrated electric field at the emitter's location. We then separately model the sensitivity of these designs to fabrication error, and incorporate this robustness as a design criteria.

\begin{figure}[!htb]
    \centering
    \includegraphics[width=0.875\linewidth]{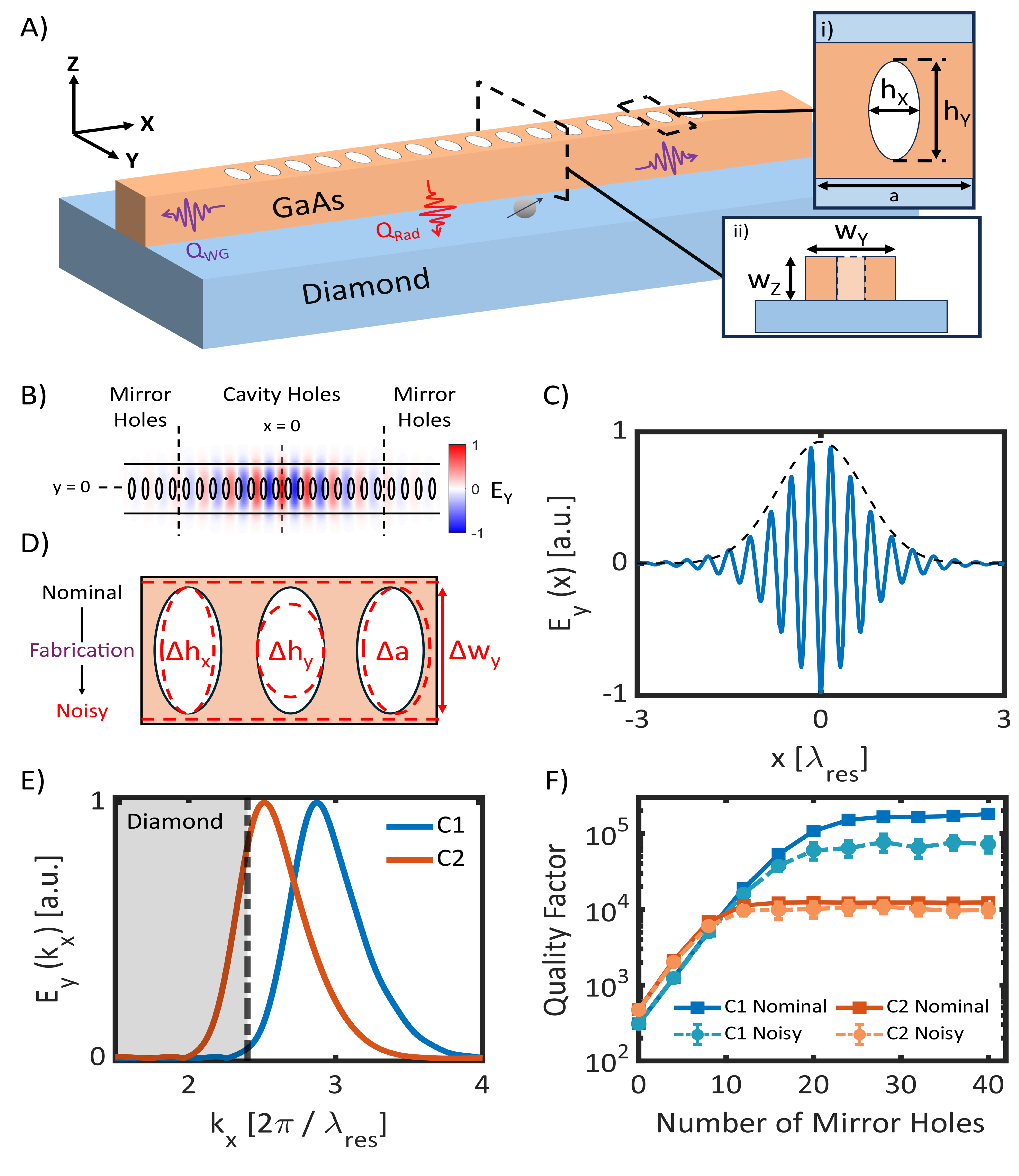}
    \caption{A) Schematic of a hybrid GaAs-on-diamond photonic crystal cavity. The cavity is defined by a one-dimensional lattice of elliptical holes with minor diameter, $h_x$, major diameter, $h_y$, lattice-constant, $a$,  beam width, $w_y$, and thickness, $w_z$. Insets i) top-down and ii) cross-sectional views of a unit-cell.
    B) 2D mode profile of a resonance overlaid with the underlying cavity geometry. The mode is concentrated at the center of the cavity and decays towards the mirror region.
    C) 1D slice of the mode profile taken at y=0. The cavity profile decays from the defect with a Gaussian envelope due to the linear increase in mirror strength from unit-cell to unit-cell in the cavity region.
    D) Schematic of fabrication-induced errors to the hole diameters $(\Delta h_x, \Delta h_y)$ and placement $(\Delta a)$, and to the beam-width $(\Delta w_y)$ of the cavity.
    E) 1D Fourier transform of the mode profiles of cavities $C_1$ and $C_2$. The shaded grey region indicates $k_x$ vectors below the diamond light line.
    F) Nominal and noisy Q-scaling of cavities $C_1$ and $C_2$. Each noisy data-point consists of 30 independent simulations. Error bars indicate 95\% confidence intervals on the fit to the mean Q. Solid and dashed-lines are guides to the eye.}
        {
    \phantomsubcaption\label{fig1a}
    \phantomsubcaption\label{fig1b}
    \phantomsubcaption\label{fig1c}
    \phantomsubcaption\label{fig1d}
    \phantomsubcaption\label{fig1e}
    \phantomsubcaption\label{fig1f}
    }
    \label{fig:fig1}
\end{figure}

We consider a one-dimensional GaAs-on-diamond photonic crystal cavity comprised of a periodic array of elliptical holes in a nanobeam waveguide as shown in figure \ref{fig1a}. The cavity unit cell is parameterized by the hole diameters, $(h_x, h_y)$, lattice constant, $(a)$, and cross-sectional area, $(w_z, w_y)$. In contrast to free-standing cavities, there are several challenges with designing hybrid photonic devices. The lack of z-symmetry complicates index guiding, and the reduced index contrast between the cavity and the substrate restricts the range of guided effective indices. Due to the high index of diamond in particular, previous designs required local etching of the diamond to achieve a large photonic bandgap \cite{barclay_hybrid_2009}. More recently, we proposed a design procedure for achieving large band-gap unit cells without etching into the diamond by performing a grid-search over the parameter space using the periodic eigenmode solver MIT photonics bands (MPB)  \cite{johnson_block-iterative_2001, huang_hybrid_2021}.

Our approach begins by sweeping the lattice constant of a unit-cell. As we are interested in modes at the $k_x = \pi/a$ point in reciprocal space, we can relate the lattice constant to a target effective index, $n_0$, as $a = \lambda_0 / 2n_{0}$ where $\lambda_0$ is the target resonance wavelength. We consider a target wavelength $\lambda_0 = 955$nm, near the zero-phonon line of the neutral SiV centre in diamond \cite{rose_observation_2018}. The effective index of a guided mode must lie between the indices of diamond and GaAs, i.e. $n_{0} \in (2.4, 3.5)$, which bounds the range of lattice-constants. For each lattice constant in this range, we independently sweep the nanobeam cross-section parameters, $w_z$ and $w_y$, and for each cross-section we vary the hole diameters according to a single parameter, the fill factor, $f \in (0,1$), such that $h_x = a\sqrt{f}$ and $h_y = w_y\sqrt{f}$. Our approach results in four independent parameters that define the design space. The metric of interest is the unit-cell mirror strength, defined as the separation between the target frequency and the nearest quasi-transverse-electric (TE) guided mode, normalized by the target frequency [S.1]. For a given unit cell, a mirror is formed by periodically arraying the cell in a one-dimensional lattice, and a cavity is created by introducing a defect in the lattice. 

In order to understand the impact of fabrication errors on cavity Q, we must first understand the origin of losses in the ideal case. It has previously been shown that the quality factor of a cavity is determined by the degree of coupling between the cavity mode and radiative modes \cite{akahane_high-q_2003}. In a hybrid geometry, the high-index substrate greatly expands the light-cone of radiative modes that can result in significant overlap between the cavity mode and leaky modes in the substrate. To achieve high-Q designs, we require a defect geometry that minimizes overlap between the cavity mode and the diamond leaky modes. We introduce a defect in the mirror lattice by quadratically chirping the lattice constant of the unit cell from the mirror periodicity, $a_{mir}$, so as to pull the fundamental quasi-TE band to create a guided mode at the target frequency [S.1]. The lattice constant that matches the fundamental TE band to the target frequency is defined as the cavity periodicity, $a_{cav}$. A quadratic chirp is chosen as it results in a linear change in mirror strength in the cavity region, yielding a Gaussian decay envelope as shown in figure \ref{fig1c} and described in \cite{quan_photonic_2010, quan_deterministic_2011}. The real-space Gaussian envelope corresponds to a Gaussian mode-profile in reciprocal space, centered at a k-vector, $k_{res} = 2 \pi n_{res} / \lambda_{res}$ where $n_{res}$ is the resonance effective index and $\lambda_{res}$ is the resonance wavelength. The adiabaticity of the chirp can be increased by using more cavity holes, resulting in higher Q at the expense of an expanded mode volume [S.2]. In this work, we consider designs with 16 cavity holes, resulting in mode volumes on the order of one cubic wavelength. 

As an example of the relationship between the cavity mode profile and Q, the reciprocal mode profile of two cavities, $C_1$ and $C_2$, are plotted in figure \ref{fig1e}, where $(a_{mir}, a_{cav}, w_z, w_y, h_x, h_y) = (172, 156, 220, 430, 71, 186)$nm for $C_1$ and $(184, 162, 220, 350, 78, 149)$nm for $C_2$. While both cavities have a similar shape to their mode profiles, $C_2$ has a lower effective index, resulting in substantial overlap with radiative modes in the diamond substrate. To illustrate the impact this overlap has on the cavity quality factors, we perform finite-difference time-domain (FDTD, Lumerical) simulations of each cavity as a function of the number of mirror holes as shown in figure \ref{fig1f}.

For a given cavity, the quality factor can be described as $1/Q = 1/Q_{wg} + 1/Q_i$ where $Q_{wg}$ refers to leakage through the ends of the cavity and into the nanobeam waveguide, and $Q_i$ represents the intrinsic cavity quality factor. As the number of mirror holes is increased, decay into the nanobeam is reduced, and the cavity quality factor approaches its intrinsic value. The intrinsic quality factor of a cavity can be approximated by the saturated quality-factor, $Q_{sat}$. The number of mirror holes required to reach saturation varies between designs according to the unit-cell mirror strength. By comparing the saturated quality factor of cavities $C_1$ and $C_2$, we observe an order of magnitude difference, consistent with the amount of spectral overlap with the leaky modes. As such, finding high-Q designs can be understood as finding designs with minimal overlap between the cavity mode and radiative modes, and thus finding designs with a high effective index for the cavity mode.

\begin{figure}[!tb]
    \centering
    \includegraphics[width=1\linewidth]{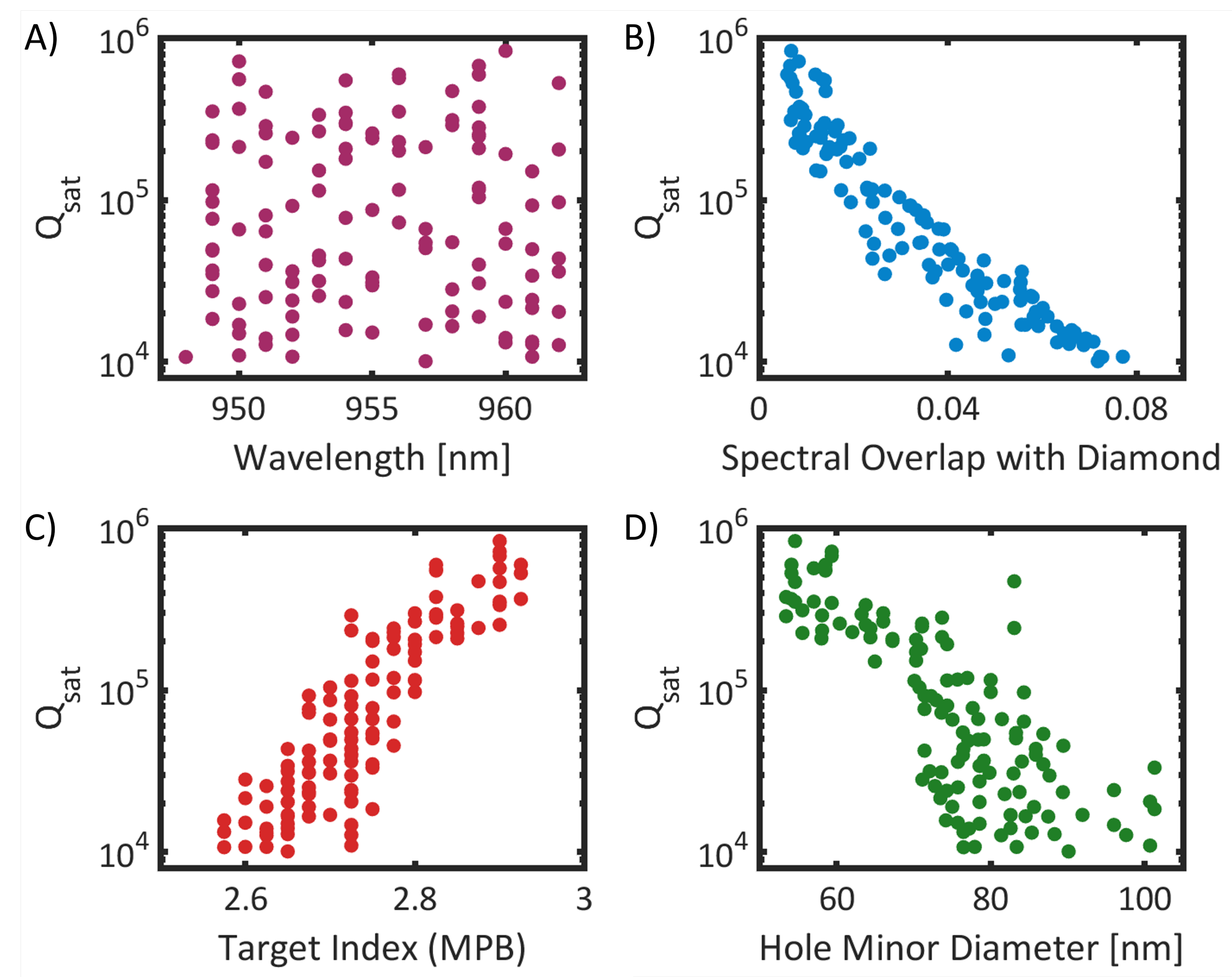}
    \caption{FDTD quality factor simulations of 200 randomly sampled cavities. Each cavity is sampled without bias from a set of 10,000 unit-cells generated from MPB simulations. 
    A) Simulated $Q_{sat}$ vs wavelength for different cavity designs near the target wavelength of 955~nm. To ensure saturation, the number of mirror holes is dynamically adjusted as a function of mirror-strength for each cavity.
    B) Saturated quality factors plotted against the cavity mode overlap with the diamond light line, 
    C) the unit-cell target index, and
    D) the hole minor-diameter, $h_x$. 
    }
    {
    \phantomsubcaption\label{fig2a}
    \phantomsubcaption\label{fig2b}
    \phantomsubcaption\label{fig2c}
    \phantomsubcaption\label{fig2d}
    }
    \label{fig:fig2}
\end{figure}

To investigate the generality of this approach, we simulate a set of 10,000 unique unit cells with appreciable mirror strength using MPB. We randomly sample from these unit cells without bias, and for each unit cell we construct a cavity by finding the appropriate $a_{cav}$. For each cavity, we simulate $Q_{sat}$ by dynamically adjusting the number of mirror holes for each design. The results for 200 such cavities near the target wavelength are shown in figure \ref{fig2a}. We observe a range of Qs from 1.0$\times 10^4$ to 8.4$\times 10^5$. By analyzing the cavity profiles, we can compare the saturated quality factors to the amount of spectral overlap with the diamond radiative modes as shown in figure \ref{fig2b}. The overlap quantity is calculated by taking the Fourier transform of a 2D slice of the cavity mode profile and computing the fraction of spectral components with $k_x$ vectors that lie below the diamond light line. We observe a clear correlation between the saturated quality factors and the amount of spectral overlap. As we use an identical chirp function for all designs, the overlap primarily depends on the resonance mode effective index [S.3]. To achieve high Q, a high effective index is thus necessary. In figure \ref{fig2c} we compare the saturated quality factor of the sampled cavities with the unit-cell target index used in the MPB sweep. As expected, we observe a clear correlation between the target effective index and the simulated quality factor. Furthermore, the mode volume can be minimized by selecting higher mirror-strength designs that compress the cavity profile along the nanobeam axis [S.3]. As such, maximizing $Q/V$ for a cavity can be simplified to maximizing the target index and mirror strength of the unit cell. By focusing on unit cell simulations, we can significantly reduce the computational overhead of the design process by deterministically screening for designs that achieve these criteria. 

While these results present a clear roadmap to designing high Q/V photonic crystals, they do not account for the practical challenges of device fabrication, where quality factors often trail the design limit by orders of magnitude \cite{wu_near-infrared_2023, ding_high-q_2024}. The discrepancies between simulated and measured quality factors can arise from a variety of fabrication imperfections such as sidewall roughness or feature placement and sizing errors. In particular, the impact of lattice disorder on light confinement has long been studied in two dimensional photonic crystal cavities and photonic crystal waveguides \cite{rodriguez_disorder-immune_2005, hughes_extrinsic_2005, hagino_effects_2009, iwaya_design_2021}. To account for such errors in the design process, we develop a simple model for simulating fabrication imperfections in our cavity geometry.

We consider errors in the hole diameters, lattice constant, and beam width of the cavity as shown in figure \ref{fig1d}. For a given cavity with some nominal saturated quality factor, $Q_0$, we generate a noisy cavity instance by applying an error to every hole and to the entire beam width. For every hole and each parameter, the errors are independently sampled from identical normal distributions with standard deviations $(\sigma_{hx}, \ \sigma_{hy}, \ \sigma_a,\  \sigma_{wy})$ respectively. 

The Q of the noisy cavity can then be directly simulated using FDTD. By repeatedly generating and simulating independent noisy cavity instances, we can extract a mean expected $Q_{sat}$ for a given cavity under a specific amount of error. To accurately estimate the fabrication errors in our devices, we image fabricated devices with scanning electron microscopy (SEM) and develop an image processing algorithm to collect statistics on the hole sizes and variance of fabricated devices [S.4]. From this analysis, we calculate an upper-bound on our errors as $(\sigma_{hx}, \ \sigma_{hy}, \ \sigma_{wy} ) = (2\%, \ 2\%, \ 1\%)$. We estimate the lattice-constant error as $\sigma_a = 0.75$~nm for our electron beam lithography tool.

\begin{figure}[!t]
    \centering
    \includegraphics[width=1\linewidth]{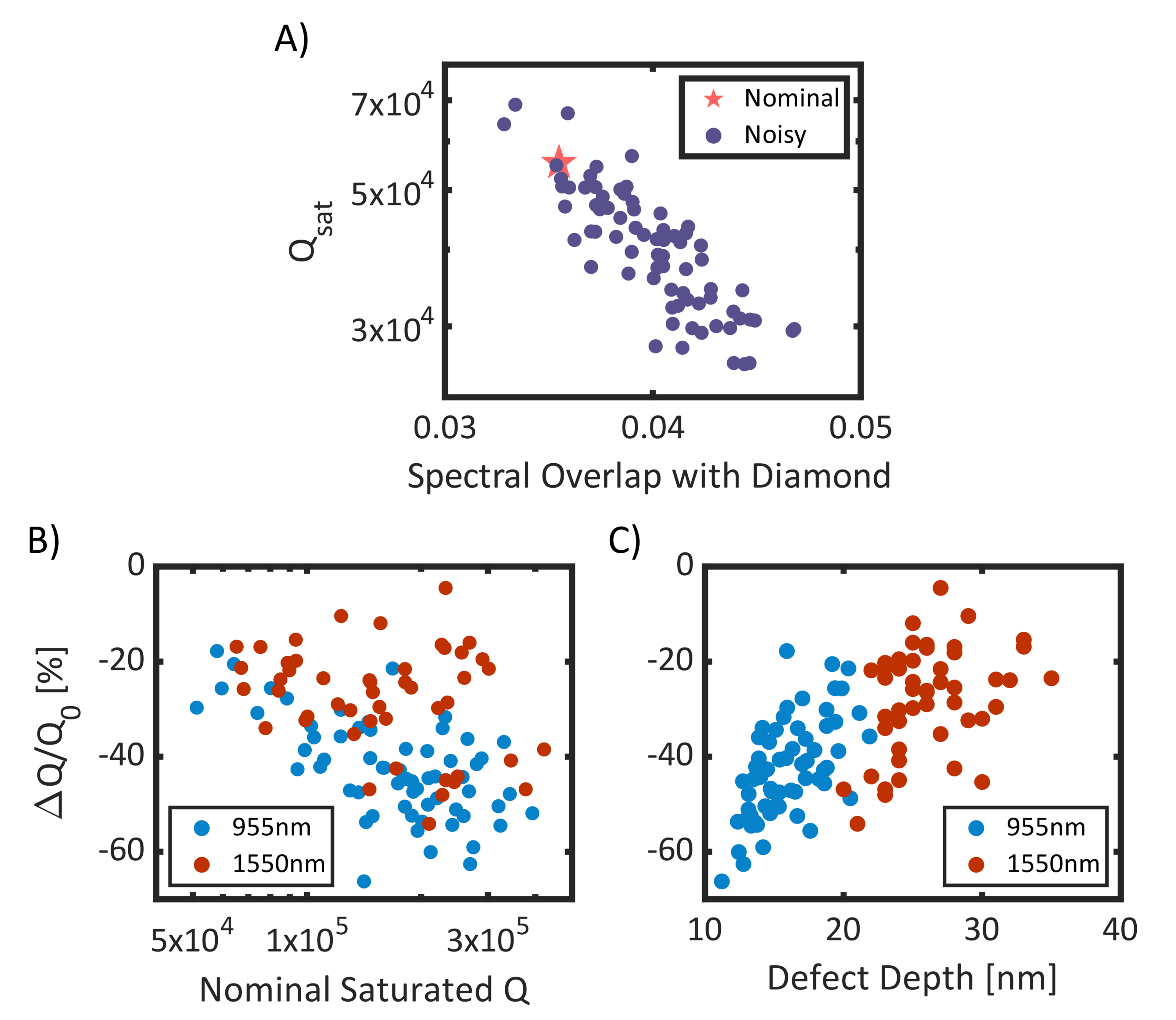}
    \caption{Fabrication error simulations. For a given cavity design, the error-sensitivity analysis is performed by repeatedly generating and simulating noisy cavity instances. The mean quality factor, ($Q_\mu$), of the noisy cavities is extracted and compared to the nominal quality factor, ($Q_0$), to compute the relative change, ($\Delta Q / Q_0$).
    A) Error-sensitivity analysis for cavity $C_3$, with $(a_{mir}, a_{cav}, w_z, w_y, h_x, h_y) = (174, 160, 220, 450, 83, 216)$nm. For each noisy-cavity simulation, the saturated quality factor and spectral profile is extracted. The decrease in quality-factor of the noisy cavities correlates with increased spectral overlap with leaky modes in  the diamond.
    B) Relative change in quality factor as a function of nominal quality factor for randomly sampled cavities at 955~nm and 1550~nm. Each datapoint is the mean of 30 noisy simulation instances.
    C) Relative change in quality factor as a function of cavity defect depth, defined as the change in period between the mirror and cavity region.
    }
        {
    \phantomsubcaption\label{fig3a}
    \phantomsubcaption\label{fig3b}
    \phantomsubcaption\label{fig3c}
    }
    \label{fig:fig3}
\end{figure}

Figure \ref{fig1f} shows an example Q-scaling simulation for cavities $C_1$ and $C_2$ in the presence of these simulated errors. At low numbers of mirror holes the nominal and noisy quality factors are similar, implying that the errors minimally affect the amount of light out-coupled to the nanobeam waveguide. Practically, this implies that the mirror strength is minimally perturbed by the errors. At larger numbers of mirror holes where Q approaches saturation, we observe a decrease in the noisy Q relative to the nominal Q. On average, the noisy cavities yield a lower intrinsic quality factor as compared to the nominal cavity. Analyzing the mode profiles of the noisy cavities at saturation, we observe that the cavities experience a perturbation to their mode profiles, which leads to increased overlap with radiative modes in the substrate as shown in figure \ref{fig3a}. 

Comparing three cavities, $C_{1,2,3}$, we observe that each design experiences a different relative change in quality factor under the same noise distributions. This indicates that certain designs are more sensitive to fabrication errors than others. The sensitivity of a cavity to fabrication errors can be calculated by the relative change in the saturated quality factor, defined as $\Delta Q / Q_0 = (Q_\mu - Q_0) / Q_0$ where $Q_\mu$ is the mean Q of the noisy cavity simulations.
To identify designs that are more robust to errors,  we randomly sample cavities from figure \ref{fig2a} and perform the noise analysis. In addition, we repeat the analysis for cavities designed to operate in the telecom band. The results show a large spread in the relative sensitivity of different designs (figure \ref{fig3b}). We observe that the telecom cavities can achieve higher robustness to fabrication errors, particularly at higher nominal Q, as compared to the 955~nm cavities. This result is consistent with prior examples in the literature where hybrid cavities in the telecom band have achieved much higher Qs than their visible wavelength counterparts \cite{chakravarthi_hybrid_2023, wu_near-infrared_2023, dibos_purcell_2022, ourari_indistinguishable_2023}.

To understand the cause of these variations, we analyze the relative change in Q as a function of various cavity parameters [S.5]. We observe a clear correlation between fabrication robustness and larger defect depths as shown in figure \ref{fig3c}. The defect depth of a cavity is defined as the total change in perodicity from the mirror region to the cavity region ($a_{mir} - a_{cav}$). This result indicates that under the applied errors, the dominant cause of increased spectral overlap is likely due to errors in the placement of the holes. Cavities with larger defect depths will necessarily require larger steps in the lattice constant for the same number of cavity holes. As such, the fractional error in placement is smaller for these cavities, resulting in a smaller relative perturbation to the mode. Designs at longer wavelengths, i.e. larger feature sizes, will necessarily experience less of a perturbation. Using this model, designs can be screened by their fabrication robustness to maximize the realized quality factor. By correlating the robustness to a single parameter, the defect depth, we can reduce the sample space when performing the noise analysis.

\section{Fabrication}
\begin{figure}[!t]
    \centering
    \includegraphics[width=1\linewidth]{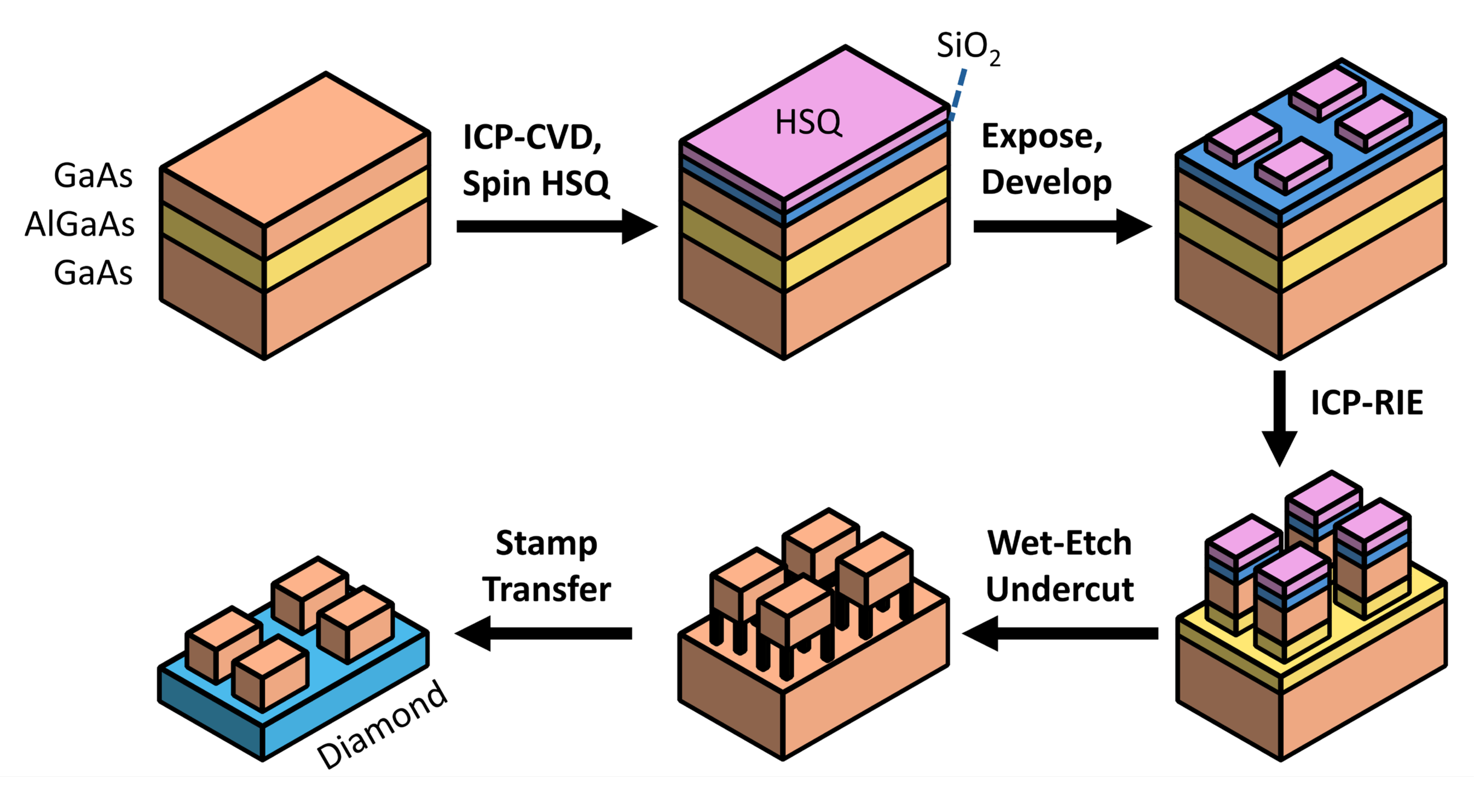}
    \caption{Fabrication flow-diagram for GaAs-on-Diamond photonic crystals. Beginning with an epitaxial GaAs-Al$_{0.8}$Ga$_{0.2}$As-GaAs wafer, we deposit a 5nm SiO$_2$ adhesion layer using an ICP-CVD, then spin HSQ and expose at 100KeV. Following development, patterns are transfered into the substrate with an ICP-RIE, then undercut and transfered to diamond for measurement.
    }
    \label{fig:fig4}
\end{figure}

We fabricate hybrid GaAs-on-diamond photonic crystals using a stamp-transfer approach \cite{dibos_atomic_2018, chakravarthi_hybrid_2023} in which devices are fabricated off-chip and then transferred onto the diamond as shown in figure \ref{fig:fig4}. Beginning with an epitaxial GaAs wafer, we pattern photonic crystals using electron-beam lithography (EBL), and transfer the patterns into the device layer using an inductively-coupled-plasma reactive ion etch (ICP-RIE). The devices are then released using a selective wet etch of a sacrificial AlGaAs layer, and transferred onto diamond using a polydimethylsiloxane (PDMS) stamp. Full fabrication details can be found in [S.6].

To achieve high Q, we focus on designs with high effective indices. In general, high effective indices will correspond to designs with larger cross-sectional areas and smaller holes. For wavelengths in the visible or near-infrared (NIR), the hole sizes required can become prohibitively small. In figure \ref{fig2d} we plot the saturated quality factor of the 955~nm cavity designs as a function of the hole minor diameter, $h_x$. We observe that for the highest Q designs, holes with $h_x$ < 60~nm are required. Writing such small holes can be challenging due to limitations in EBL resist contrast and resolution. Furthermore, etching small holes or trenches incurs effects of aspect-ratio dependent etching (ARDE) \cite{lee_featuresize_1991}. While smaller features can be written by using thinner resist, this can lead to mask erosion and roughness during the etching process. This is further exacerbated by the ARDE phenomenon as longer etches are required to fully clear the small features. 

In order to verify the accuracy of the fabrication sensitivity simulations of the previous section, it is critical that device performance is not limited by poor fabrication in the form of sidewall roughness, imperfect hole etching, or residue. As such, we develop a fabrication procedure for photonic crystals with small holes by optimizing the EBL, etching, and undercut processes. To write small holes in a device, it is necessary to use a high-contrast, high-resolution resist with excellent etch selectivity. We use hydrogen silsesquioxane (HSQ) as it exhibits high selectivity which allows for thinner resists and thus smaller features. However, HSQ contrast is much lower than other common EBL resists such as polymethyl methacrylate (PMMA) or ZEP \cite{chen_nanofabrication_2015}. As such, there has been significant effort to maximize the HSQ contrast by manipulating the development conditions through post-exposure baking \cite{kim_nanometer-scale_2013} and developing at elevated temperatures \cite{chen_effects_2006}. 

\begin{figure}[!t]
    \centering
    \includegraphics[width=1\linewidth]{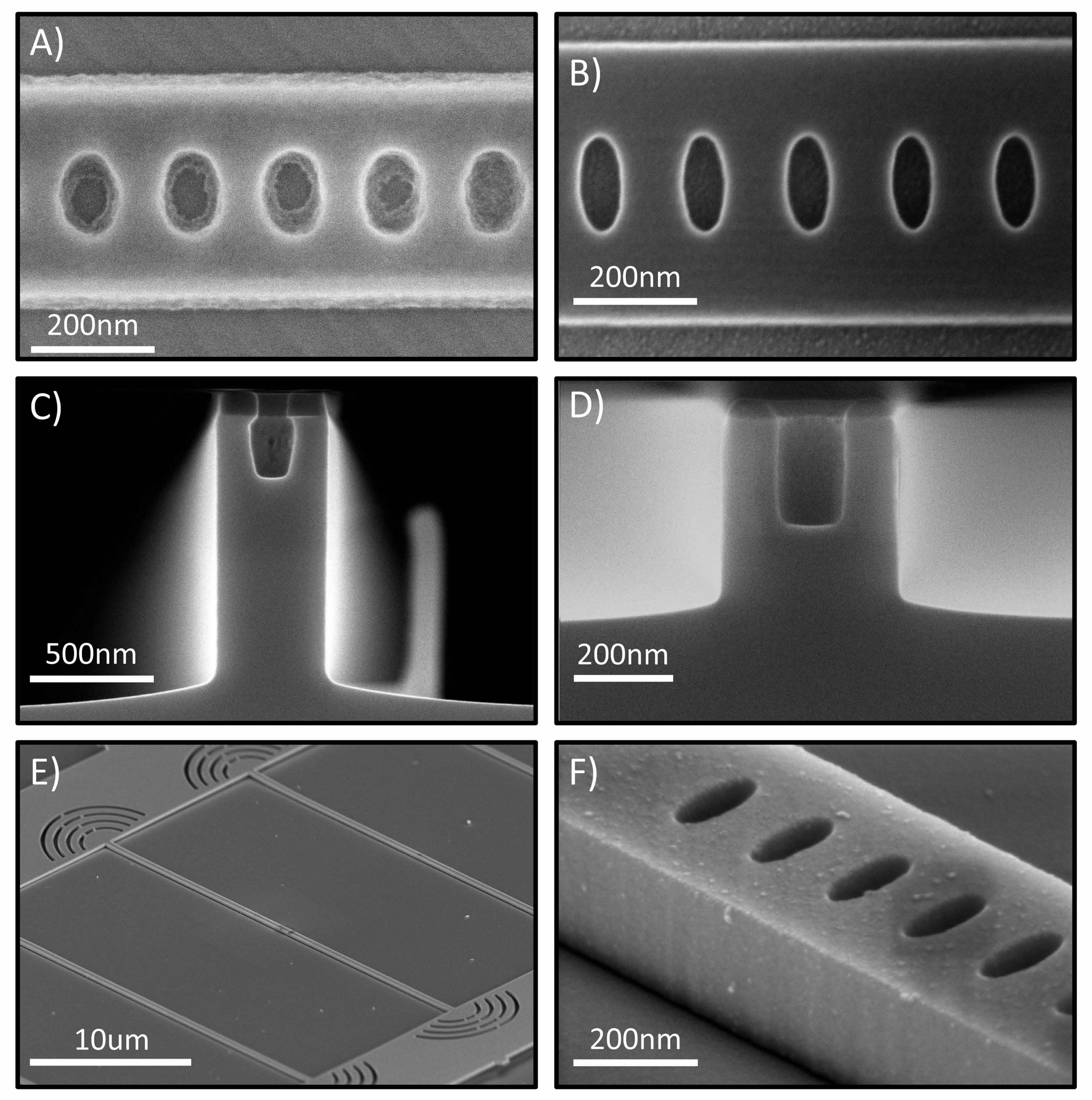}
    \caption{
    A) Example photonic crystal written in HSQ and developed using standard TMAH. The pattern was written at a dose of 650 uc/cm2 and developed in 25\% TMAH for 2 minutes.
    B) Example photonic crystal written in HSQ and developed using salty-TMAH. Sample was written at a dose of 5700 uc/cm2 and developed in 1\% NaCl in 25\% TMAH at 30$^{\circ}$C for 4 minutes.
    C) Cross-sectional SEM image of a photonic crystal-hole with ($h_x$ = 70nm, $h_y$ = 130nm) etched with a high-flow etch. The etch was performed with 10sccm each of Cl$_2$, BCl$_3$, Ar$_2$, and N$_2$ at a pressure of 5mTorr, RF Power of 50W and ICP of 500W. Etch time was 30s.
    D) Cross-sectional SEM image of the ARDE optimized etch. The etch was performed with 1.75sccm Cl$_2$, 1sccm N$_2$, and 2sccm Ar at a pressure of 2mTorr, RF power of 25W and ICP of 50W. Etch time was 330s.
    E) SEM image of a device pattern after stamp-transfer on diamond. The sample was coated with 3nm of iridium to prevent charging during imaging. 
    F) Magnified image of the photonic crystal nanobeam. The observed residue on the cavity results from the iridium sputtering process for imaging.
    }
        {
    \phantomsubcaption\label{fig5a}
    \phantomsubcaption\label{fig5b}
    \phantomsubcaption\label{fig5c}
    \phantomsubcaption\label{fig5d}
    \phantomsubcaption\label{fig5e}
    \phantomsubcaption\label{fig5f}
    }
    \label{fig:fig5}
\end{figure}

In figure \ref{fig5a} we show an example of a section of a photonic crystal waveguide written in HSQ and developed in 25\% tetramethylammonium hydroxide (TMAH) in water. The sample is baked before and after development at 200$^{\circ}$C on a hotplate. For a given design, as the written dose increases, the holes approach the target dimensions before saturating. Further increasing the dose, or reducing the pattern diameters, results in undesired HSQ development within the holes. By manipulating the density of nearby features, the degree of overdosing within the holes can be reduced, indicating that electron scattering from nearby writing is a key factor limiting the hole contrast [S.6.1]. Under these development conditions, we observe a minimum hole diameter of $h_x =$ 85nm. 

To write the holes necessary for the highest Q designs, we seek to further optimize the development process to improve the HSQ contrast without sacrificing resist thickness or pattern density. It has previously been observed that the addition of salts to alkali developers can significantly increase the contrast of HSQ by enhancing the dissolution rate, leading to a more aggressive development \cite{yang_using_2007, kim_understanding_2009}. The contrast has been shown to further be enhanced by optimizing the temperature of the solution during development \cite{yan_effects_2010}. While these results have primarily focused on maximizing the density and size of lines, we expect the increased contrast to aid in mitigating the effects of electron scattering in the holes of our photonic crystals. We prepare a mixture of 1\% NaCl in 25\% TMAH, heated to 30$^{\circ}$C, hereto referred to as ``salty TMAH". In comparison to standard TMAH development ($\gamma = 2.03$), we observe a significant increase in contrast ($\gamma = 3.83$) for the salty TMAH [S.6.1]. The saturated dose increases from 1150~uC/cm$^2$ for standard TMAH to 4900~uC/cm$^2$ for salty TMAH. In figure \ref{fig5b} we show an example of a  pattern developed in salty TMAH. The mask achieves significantly higher contrast and reduced line-edge roughness, and minor diameters as small as 55~nm can be written with a high-density of nearby writing. One trade-off with the salty development is the high dose required, which significantly increases the required write time. However, higher doses allow for improved etch selectivity as the HSQ matrix density is increased by the longer exposure \cite{yang_enhancing_2006}.

While these results allow for writing small holes, an additional challenge is to transfer the patterns into the device layer via etching. In a single-step etch process, pattern transfer is achieved through a combination of simultaneous physical etching, chemical etching, and passivation. As an example, in figure \ref{fig5c} we show a cross-sectional SEM image of a typical hole in a photonic crystal device. The pattern was etched using a combination of Cl$_2$, Ar, BCl$_3$, and N$_2$ gases at high flow rates and high power. After etching, the chip was scribed and manually cleaved before mounting vertically within an SEM to image the cross section. In this etch, the chlorine radicals generated in the plasma react with the GaAs substrate through a chemical process. The chemical reaction between the chlorine and GaAs is isotropic in nature, and so to achieve vertical sidewalls high RF powers are used to direct ICP radicals. This however results in significant physical bombardment of the mask, reducing the achievable selectivity. To overcome this challenge, passivating gases are used to help achieve anisotropic etching with more moderate RF powers \cite{franz_high-rate_1998, franz_sidewall_2001, edwards_fabrication_2007}. In particular, N$_2$ has been shown to be an excellent passivating agent for Cl$_2$ based etching of GaAs \cite{maeda_inductively_1999, volatier_extremely_2010}. High flow rates are typically used for the chemical gases to achieve rapid etch rates in the substrate while minimizing mask erosion. By tuning these parameters, it is possible to achieve extremely high selectivity and vertical sidewalls in macroscopic regions. 

However, etching inside the small holes of the photonic crystal introduces several challenges. The etch rate is severely reduced inside the hole, and also exhibits a high degree of isotropy, leading to a "bottling" effect. The reduction of the etch rate as a function of feature size is known as RIE-lag \cite{lee_featuresize_1991} and is a form of ARDE. ARDE is a phenomenon in which the etch rate within a hole or trench decreases as the aspect ratio increases. The aspect ratio is defined as the etch depth divided by the width of the opening. As the etch depth increases the aspect ratio also increases, further reducing the etch rate and leading to a self-limiting effect \cite{gottscho_microscopic_1992}. As a result, using thicker masks effectively increases the aspect ratio of the structure, and thus ARDE cannot be overcome by simply using thicker masks with longer etch times. The microscopic origin of ARDE can be explained by neutral shadowing \cite{bailey_scaling_1995}, while the increased isotropy can be explained by radical scattering within the holes \cite{shaqfeh_simulation_1989}. 

To overcome these challenges, we develop an etch recipe optimized for hole-based photonic crystal cavities by simultaneously minimizing ARDE, maximizing selectivity, and minimizing sidewall roughness. To mitigate the effects of radical scattering, we seek to minimize the overall density of radicals within the plasma. We begin by reducing the gas flow rates to the minimum supported by the tool, and reduce the ICP power to the minimum able to support a stable inductively-coupled plasma. To improve selectivity, we reduce physical bombardment of the mask by reducing the RF power to the lowest stable value. The pressure of the chamber is reduced to the lowest supported value to reduce sidewall roughness, while the Ar flow is adjusted to stabilize the plasma. Finally, the relative ratios of Cl$_2$ and N$_2$ are adjusted to deterministically control the sidewall angle of the holes and beam to achieve vertical profiles [S.6.2]. The resulting etch is shown in figure \ref{fig5d} and demonstrates a significant reduction in ARDE while maintaining vertical sidewalls. 

Following etching, the sacrificial layer is partially removed using hydrochloric acid, and the HSQ is stripped using dilute hydrofluoric acid. The undercut parameters are selected to fully suspend the cavities while mitigating residue [S.6.3]. After the undercut, the devices are transferred onto diamond using a PDMS stamp. In figure \ref{fig5f} we show an example device after stamp-transfer onto diamond. The resulting devices exhibit smooth, vertical sidewalls and minimal residue. 

\section{Measurement}

\begin{figure}[!t]
    \centering
    \includegraphics[width=1\linewidth]{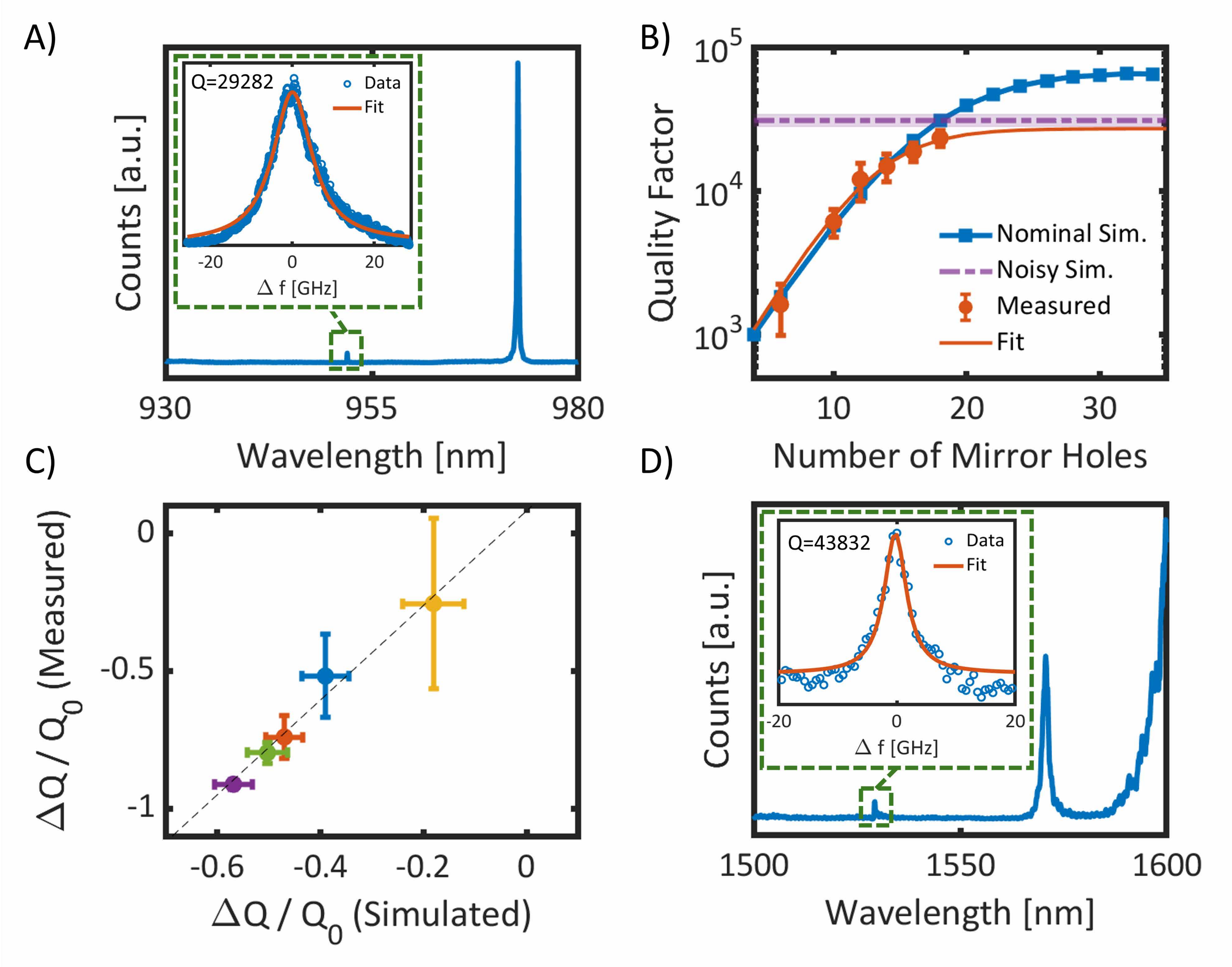}
    \caption{
    A) Transmission spectrum of a typical cavity. The resonance at 952nm is the fundamental mode, while the feature at 975nm is a higher-order mode. Inset: Lorentzian fit of the resonance from which the quality factor is extracted.
    B) Q-scaling measurements of a cavity design compared to simulation. The blue data points are the simulated Q-scaling of the cavity in the absence of fabrication errors. The blue solid line is a guide to the eye. The purple dashed-curve is the mean-saturated quality factor of 70 noisy cavity simulations, and the shaded purple region represents 95\% confidence intervals on the fit to the mean. Red data-points are the mean and standard deviation of the measured quality factors, and the solid red line is a fit to the measurement data.
    C) Relative change in quality factor of five fabricated cavity designs compared to simulation. The dashed line is a linear fit to the data. X-error bars are 95\% confidence intervals on the fit to the mean noisy-Q.  Y-error bars are 95\% confidence intervals on the fitted intrinsic Q.
    D) Example transmission measurement of a cavity in the telecom C-band.
    }
    {
    \phantomsubcaption\label{fig6a}
    \phantomsubcaption\label{fig6b}
    \phantomsubcaption\label{fig6c}
    \phantomsubcaption\label{fig6d}
    }
    \label{fig:fig6}
\end{figure}

To verify our fabrication error modelling, we fabricate several different cavity designs with a similar nominal quality factor, but with a range of different simulated sensitivities to fabrication errors. Figure \ref{fig5e} shows an example of a pattern of devices on diamond. A device consists of a single cavity with grating couplers on both ends, and each pattern consists of seven identical cavities. For a given cavity design, we sweep the number of mirror holes across different patterns. The entire set of device patterns is then repeated with shifts to the hole sizes and beam widths to account for write-to-write variations, resulting in 210 devices for each cavity design.

Transmission measurements of the cavities are performed using a free-space confocal microscopy setup [S.7]. Figure \ref{fig6a} shows the transmission spectrum of a typical cavity. By measuring cavities at different numbers of mirror holes, we can fit the intrinsic quality factor of the fabricated cavity. The measurements are fit to a model of $1/Q = 1/Q_{wg} + 1/Q_i$ with $Q_{wg} = Ae^{bM}$ where $M$ is the number of mirror holes. The fitting parameters $(A, b, Q_i)$ are left as free parameters without bounds. In physical terms, $A \equiv Q(M=0)$ is the quality factor of the cavity at zero mirror holes. The parameter $b$ is related to the mirror strength of the design and dictates the rate at which the cavity reaches saturation. Figure \ref{fig6b} shows a typical Q-scaling fit of a cavity design, showing good agreement between the simulated saturated Q in the presence of fabrication error, and the fitted intrinsic Q. 

We repeat this process for five different cavity designs with different simulated robustness. The full measurements can be found in [S.8]. By comparing the fitted intrinsic Q to the predicted noisy saturated Q, we find excellent agreement in the relative sensitivities to fabrication error of the different designs. These results indicate that e-beam placement and feature size-errors are indeed the dominant loss mechanism for the cavities. Additionally, we observe that the fabricated devices demonstrate a higher degree of sensitivity to errors than the simulated values, further illustrating the importance of selecting robust designs.

To illustrate the generality of our design and fabrication process, we fabricate photonic crystals designed for the telecom C-band. In figure \ref{fig6d}, we demonstrate an example transmission measurement of a cavity with a resonance near 1520~nm with quality factor approaching 45,000. This result is comparable to recent demonstrations of hybrid silicon photonic crystal cavities with comparable index contrast \cite{dibos_atomic_2018, dibos_purcell_2022, yu_frequency_2023, huang_stark_2023}.

\section{Conclusion}
In this work, we demonstrate that robustness to fabrication errors is a critical design parameter for hybrid photonic crystal cavities that can be directly modelled in FDTD. The simulated robustness varies across designs, independent of the nominal quality factor, which is the typical parameter that is optimized in the design process. 

We show that the nominal Q is determined by the degree of coupling between the cavity mode and substrate leaky modes, and can be maximized by selecting high effective index unit cells. By correlating the nominal Q, mode volume, and robustness to unit cell parameters, we are able to screen unit cells based on their potential to maximize $Q/V$ and fabrication robustness. This result significantly reduces the computational burden of the design process, as specific unit cells can be deterministically selected for simulation rather than relying on random sampling. As the unit cell simulations can be performed rapidly in comparison to full cavity FDTD simulations, we can probe a significant portion of the design space with reduced computational overhead. Additionally, optimization techniques could be used to create unit cells that simultaneously achieve the desired attributes.

Using our fabrication sensitivity model, we can search for designs which simultaneously maximize the nominal quality factor while minimizing fabrication error sensitivity. We fabricate cavities with target $Q_0 > 1 \times 10^5$ at 955~nm, but we observe a departure from the model in fabricated devices at high Q around 30,000, independent of simulated robustness or fabrication processes [S.9]. There are many factors that could limit the device quality factor, including scattering losses related to fabrication or material absorption. As we observe a similar limit at both 955~nm and 1520~nm, we can exclude fabrication-related scattering as this would scale as $\lambda^{-4}$. Similarly, multi-photon absorption in the material would be expected to scale more sharply with wavelength. As such, we hypothesize that the observed Q arises from material absorption due to bulk or surface defects. In particular, we observe that using commercial metalorganic chemical vapour deposition (MOCVD) GaAs wafers, the device quality factors are limited to Q below 10,000 at 955~nm [S.10]. By instead using high purity molecular-beam epitaxy (MBE) wafers, we observe an immediate threefold increase in quality factor for the high-Q designs. Owing to the ultra-clean vacuum environment and novel chamber design, these films have achieved record mobility with implied bulk defects below 1 part in 10 billion \cite{chung_ultra-high-quality_2021, chung_understanding_2022, gupta_ultraclean_2024}. As such, we conclude that the improvement in device Q from the MOCVD to MBE wafers is likely due to a reduction in bulk defects. 

One potential source of loss may be mid-gap states in the surface oxide, specifically along the device sidewalls. Previous work has shown that surface passivation techniques can dramatically improve device Q in GaAs photonic devices \cite{parrain_origin_2015, guha_surface-enhanced_2017, stanton_efficient_2020, kuruma_surface-passivated_2020, thomas_quantifying_2023}. Alternatively, it has also been shown that oxygen segregation during growth of the sacrificial AlGaAs layer can create impurities that propagate into the device layer \cite{chung_surface_2018}. As we use a large sacrificial layer thickness for device suspension and transfer (see S.6.3) the quantity of oxygen-segregated defects may be significant in our device layer. To further improve device Q, modifications to the wafer stack, or the addition of passivation steps may be required. Despite these limitations, the demonstrated cavity quality factors at 955~nm represent the highest experimentally realized Q for hybrid cavities in the visible or NIR to the best of our knowledge, while our results at 1520~nm are comparable to the state of the art for silicon cavities with similar index contrast. These results were enabled by improvements in the design procedure, as well as through optimization of the fabrication processes. 

Owing to the generality of the design process, our model could be readily applied to arbitrary wavelengths and material stacks. By separating device fabrication from the substrate, the hybrid photonics platform could be utilized to significantly expand the space of candidate qubits for quantum network experiments. The full design code used in this work is available at \href{https://github.com/deleon-photonics/Photonics_Simulations}{github/deLeonPhotonics}. 
\vspace{0.5cm}

\textit{Acknowledgements: }We thank Mouktik Raha, Jeff Thompson, and Ding Huang for help with the initial MPB code used in this work. The simulations presented in this article were performed in part on computational resources managed and supported by Princeton Research Computing, a consortium of groups including the Princeton Institute for Computational Science and Engineering (PICSciE) and the Office of Information Technology's High Performance Computing Center and Visualization Laboratory at Princeton University. The authors acknowledge the use of Princeton's Imaging and Analysis Center (IAC), which is partially supported by the Princeton Center for Complex Materials (PCCM), a National Science Foundation (NSF) Materials Research Science and Engineering Center (MRSEC; DMR-2011750). The devices in this work were fabricated at the Princeton Materials Institute Micro/Nanofabrication Center.

\newpage
\bibliographystyle{ieeetr} 
\bibliography{main}

\end{document}


\maketitle

\section{Cavity Design and Photonic Band Diagrams}
\label{S.1}

\begin{figure}[h!tbp]
    \centering
    \includegraphics[width=1\linewidth]{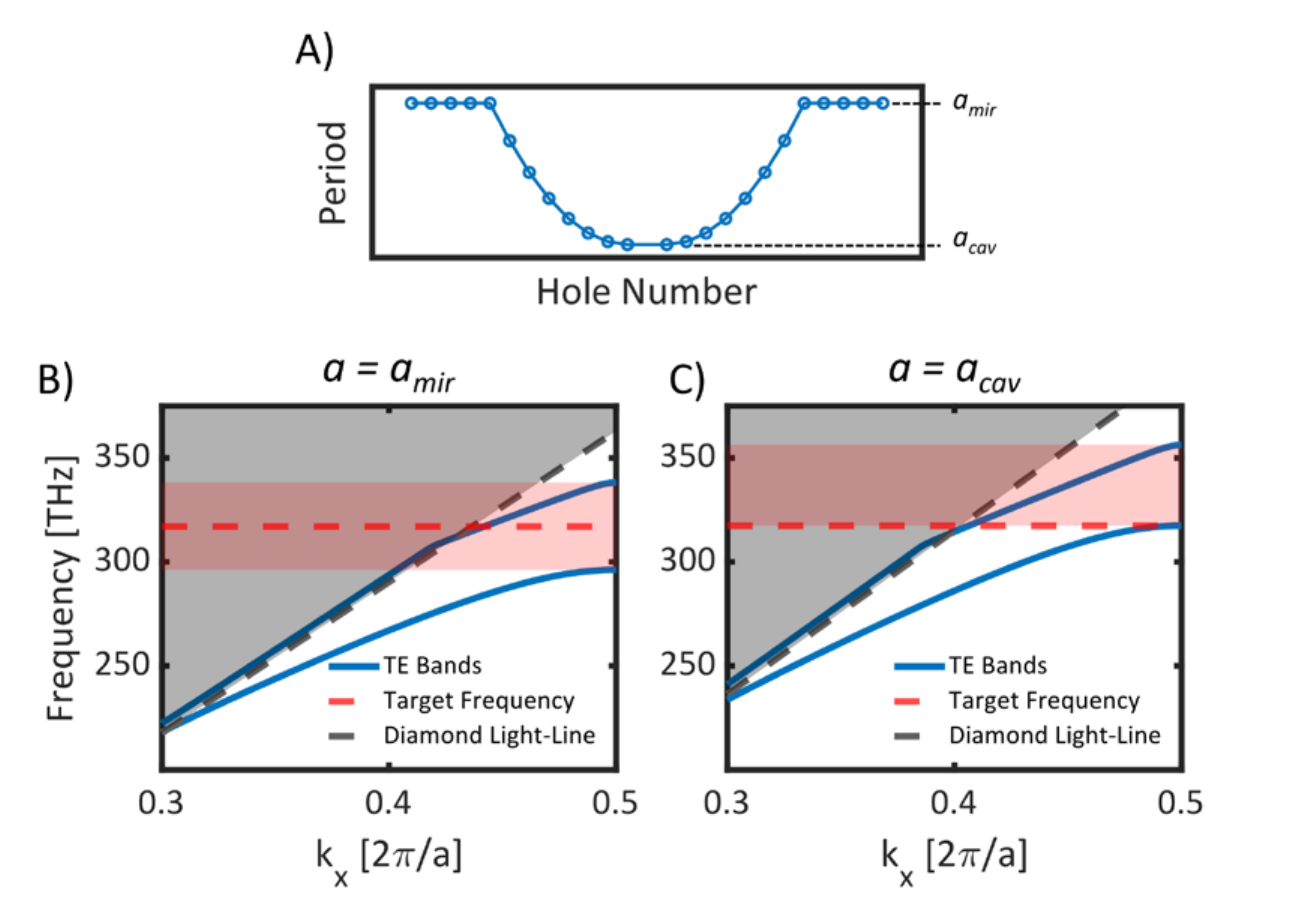}
    \caption{Cavity construction and band diagrams. 
    A) Cavity periodicity as a function of hole number. The period of the cavity holes are chirped from unit cell to unit cell using a quadratic taper function. 
    B) Example band diagram of a mirror hole for cavitiy $C_1$. The shaded red-region indicates the band-gap for quasi TE modes
    C) Example band diagram at the center of the cavity. The quasi TE bands are pulled to higher frequencies by decreasing the period until the fundamental band matches the target frequency.}
    {
    \phantomsubcaption\label{SI1a}
    \phantomsubcaption\label{SI1b}
    \phantomsubcaption\label{SI1c}
    }
    \label{fig:SI_1}
\end{figure}

In figure \ref{fig:SI_1} we outline the procedure for designing the one-dimensional photonic crystal cavities discussed in this work. As discussed in the main text and in a prior work \cite{huang_hybrid_2021}, we begin by analyzing single unit cells consisting of an elliptical hole in a nanobeam waveguide. Using MIT photonics bands (MPB) \cite{johnson_block-iterative_2001}, we simulate the frequency bands of a unit cell at a given point in reciprocal space. The metric of interest is the mirror strength, defined according to equation \ref{SI_eqn_MirStr}, where $f_0, f_{TE0}$ and $f_{TE1}$ are the target frequency and frequencies of the first two quasi TE bands at the $k_x = \pi/a$ point in reciprocal space.

\begin{equation}
\rm{Mirror \ Strength} \equiv \frac{min\left[ (f_{TE1} - f_0), (f_0 - f_{TE0}) \right]}{f_0}
\label{SI_eqn_MirStr}
\end{equation}

By varying the cross-sectional area, periodicity, and fill factor, we can perform a grid search of possible unit cells. In figure \ref{SI1b} we plot the full quasi TE band diagram of the mirror unit-cell for cavity $C_1$ from the main text. For the emitter to couple to the cavity, it is is necessary to shift the bands such that a guided mode exists at the target frequency. This can be achieved by modifying the periodicity of the unit-cell while keeping all other parameters fixed. As shown in figure \ref{SI1c}, reducing the periodicity of the hole shifts the quasi TE modes to higher frequencies. The number of cavity holes and the functional form of the chirp from the mirror region to the cavity region dictates the adiabaticity of the introduced defect mode. We use a quadratic chirp function where the periodicity of a given unit-cell in the cavity region is given by 

\begin{equation}
a(i) = a_{cav} + \frac{(a_{mir}-a_{cav})}{N^2} i^2
\label{SI_eqn_QuadraticChirp}
\end{equation}
where $N = \frac{N_{cav}}{2} - 1$ and $i \in (0, 1, 2, ..., N)$. $N_{cav}$ is the total number of cavity holes used for the chirp. The functional form is chosen so that the first hole in the cavity region is at the cavity periodicity, i.e. $a(0) = a_{cav}$, and the last hole is at the mirror periodicity, $a(N) = a_{mir}$ as shown in figure \ref{SI1a}.

\newpage
\section{Cavity Chirp Adiabaticity}
\label{S.2}

\begin{figure}[h!tbp]
    \centering
    \includegraphics[width=1\linewidth]{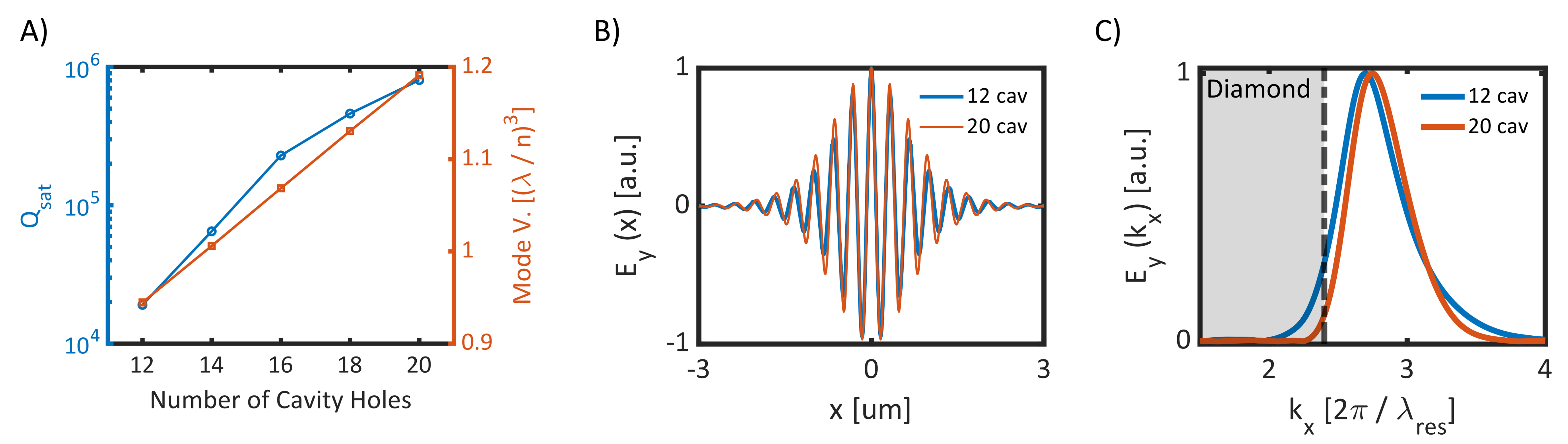}
    \caption{
    A) Saturated quality factor and mode-volume of cavity $C_1$ as a function of the number of cavity holes. 
    B) Real-space mode-profile of the cavity mode with 12 and 20 cavity holes.
    C) Fourier transform of the real-space mode profiles normalized by the simulated resonance wavelength.}
    {
    \phantomsubcaption\label{SI2a}
    \phantomsubcaption\label{SI2b}
    \phantomsubcaption\label{SI2c}
    }
    \label{fig:SI_2}
\end{figure}

To understand the impact of the number of cavity holes in the chirp region, we consider the example of cavity $C_1$. We simulate the saturated quality factor (Q$_{sat}$) and mode profile of the cavity as a function of the number of cavity holes as shown in figure \ref{fig:SI_2}. As the number of cavity holes is increased, Q$_{sat}$ increases, but so does the mode-volume. As the metric of interest is $Q/V$, choosing the exact number of cavity holes can depend on the Q-factors achievable in practice for a given material and fabrication process. The increase in both mode volume and quality factor can be explained by studying the cavity mode profiles. As an example, in figure \ref{SI2b} we compare the real-space mode profiles along the nanobeam axis for cavity $C_1$ with 12 and 20 cavity holes. We observe that for 12 cavity holes, the mode-profile decays more rapidly along the nanobeam axis. As we are using fewer holes to chirp from $a_{cav}$ to $a_{mir}$, the mirror-strength is increased in larger steps as compared to 20 cavity holes. This narrowing of the mode-profile in real-space necessarily corresponds to a broadening in reciprocal space, and thus greater spectral overlap with diamond leaky modes as shown in figure \ref{SI2c}. 

\section{Additional Unit Cell Simulation Results}
\label{S.3}

\begin{figure}[h!tbp]
    \centering
    \includegraphics[width=1\linewidth]{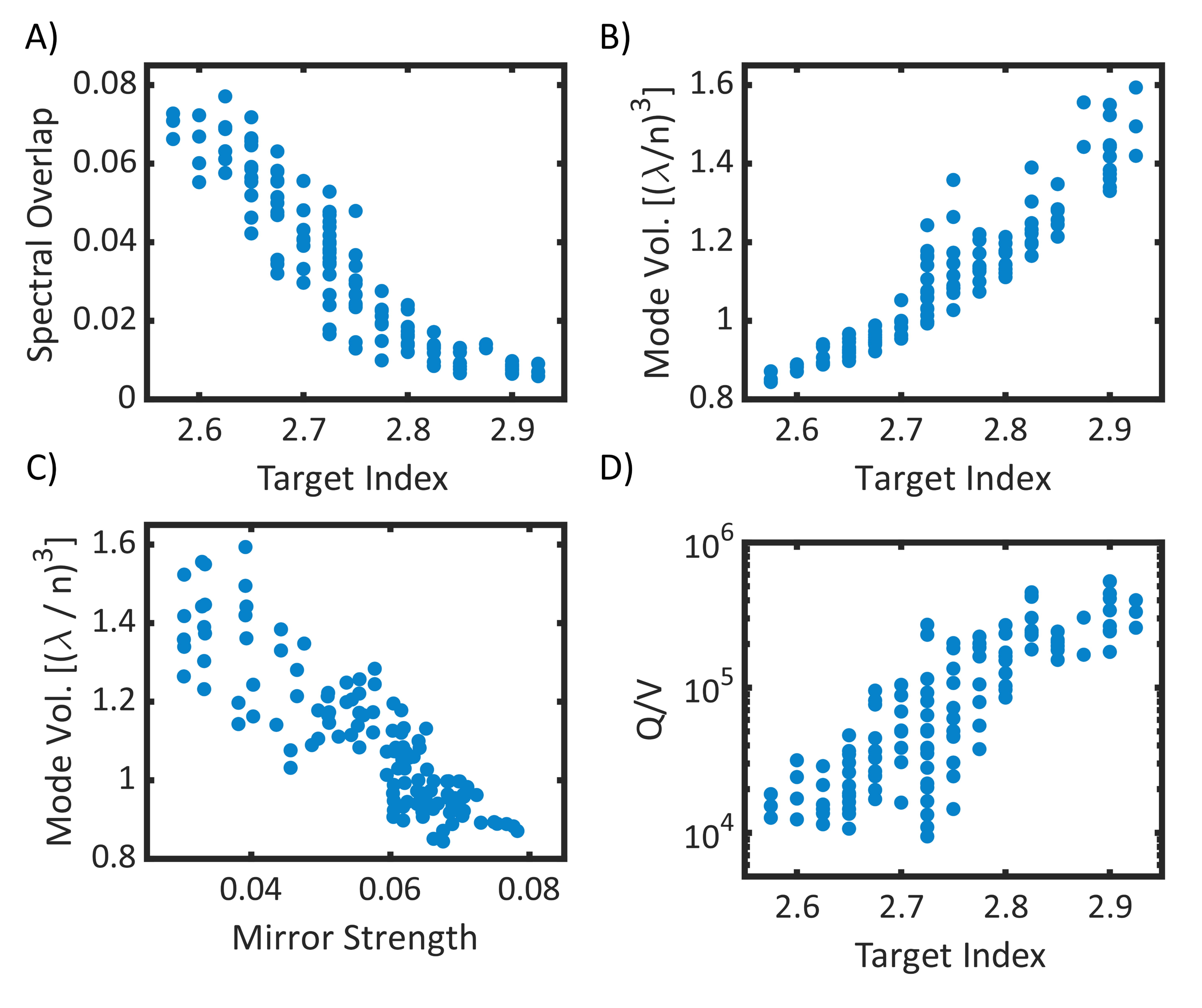}
    \caption{
    A) Spectral overlap between the cavity mode-profiles and the diamond leaky-modes plotted as a function of the MPB target index.
    B) Cavity mode volume as a function of the target-index.
    C) Cavity mode volume as a function of the unit-cell mirror strength.
    D) Q/V as a function of target index.}
    {
    \phantomsubcaption\label{SI3a}
    \phantomsubcaption\label{SI3b}
    \phantomsubcaption\label{SI3c}
    \phantomsubcaption\label{SI3d}
    }
    \label{fig:SI_3}
\end{figure}

In figure 2 of the main-text, the Q$_{sat}$ of 200 cavity designs and their mode-profiles are analyzed. We observe that the saturated quality factor depends on the amount of spectral overlap with leaky modes in the diamond. As we use an identical quadratic chirp function for all cavities, we expect that the amount of spectral overlap will primarily depend on the effective index of the cavity resonance. In figure \ref{SI3a} we plot the amount of spectral overlap between the cavity mode and the diamond leaky modes as a function of the MPB target index. We observe a clear anti-correlation between the target indices and the spectral overlap. As such, higher target indices result in less overlap. However, we observe a trade-off in mode-volume as shown in figure \ref{SI3b}. Despite the apparent trade-off, the ratio of Q/V for the cavities still scales with the target index, illustrating the importance of using high effective index designs. Additionally, we observe that the mode-volume of the cavity depends on the unit-cell mirror strength as shown in figure \ref{SI3c}. Designs with larger mirror-strength will necessarily yield a more compressed mode-profile along the nanobeam axis. This points to two parameters, $n_0$ and mirror-strength, that can be optimized simultaneously at a unit-cell level to achieve high Q/V cavities.

\section{Fabrication-Error Analysis}
\label{S.4}

\begin{figure}[h!tbp]
    \centering
    \includegraphics[width=1\linewidth]{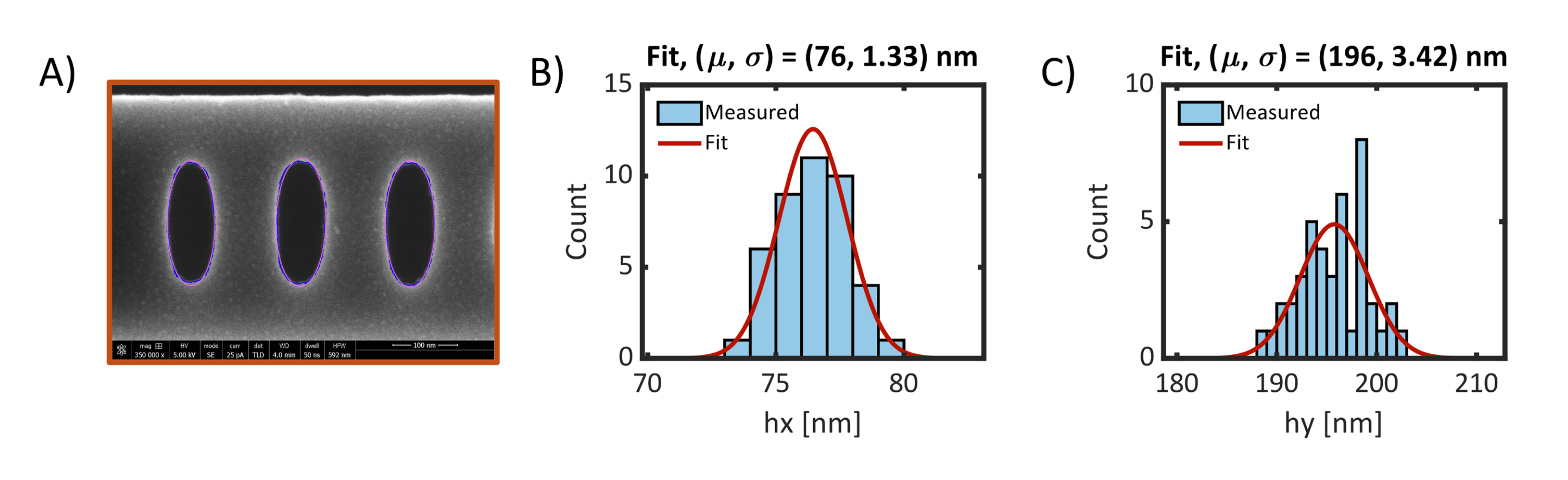}
    \caption {
    A) Example SEM image of the holes in a fabricated cavity. The blue data-points are the identified contour edges, and the purple ellipses indicate the fit to the contours. Images are taken at 5KeV and 25pA at a magnification of 350,000x (Thermo-Scientific Verios-460 XHR SEM).
    B) $h_x$ and C) $h_y$ fit results of the holes in the device.
    }
    {
    \phantomsubcaption\label{SI4a}
    \phantomsubcaption\label{SI4b}
    \phantomsubcaption\label{SI4c}
    }
    \label{fig:SI_4}
\end{figure}

To estimate the amount of error to use in our robustness simulations, we develop an image-processing algorithm to directly analyze scanning electron microscope (SEM) images of our fabricated devices. In figure \ref{fig:SI_4} we show the results of an example fabrication run and subsequent analysis. Following electron beam lithography, etching, and undercutting of the cavity design (see section \ref{S.6}), we acquire SEM images of the holes in the the fabricated cavities. For a given number of cavities across the chip, we image every hole in the nanobeam. The algorithm uses the openCV package \cite{bradski_opencv_2000} to identify the hole contours and fit them to ellipses as shown in figure \ref{SI4a}. From this fit, we extract the minor and major diameters of the holes. In figures \ref{SI4b} and \ref{SI4c} we show the results for an example cavity. By analyzing the distribution across the holes for different designs, we estimate an upper bound on the hole errors.

\section{Additional Fabrication Robustness Correlations}
\label{S.5}
\begin{figure}[h!tbp]
    \centering
    \includegraphics[width=1\linewidth]{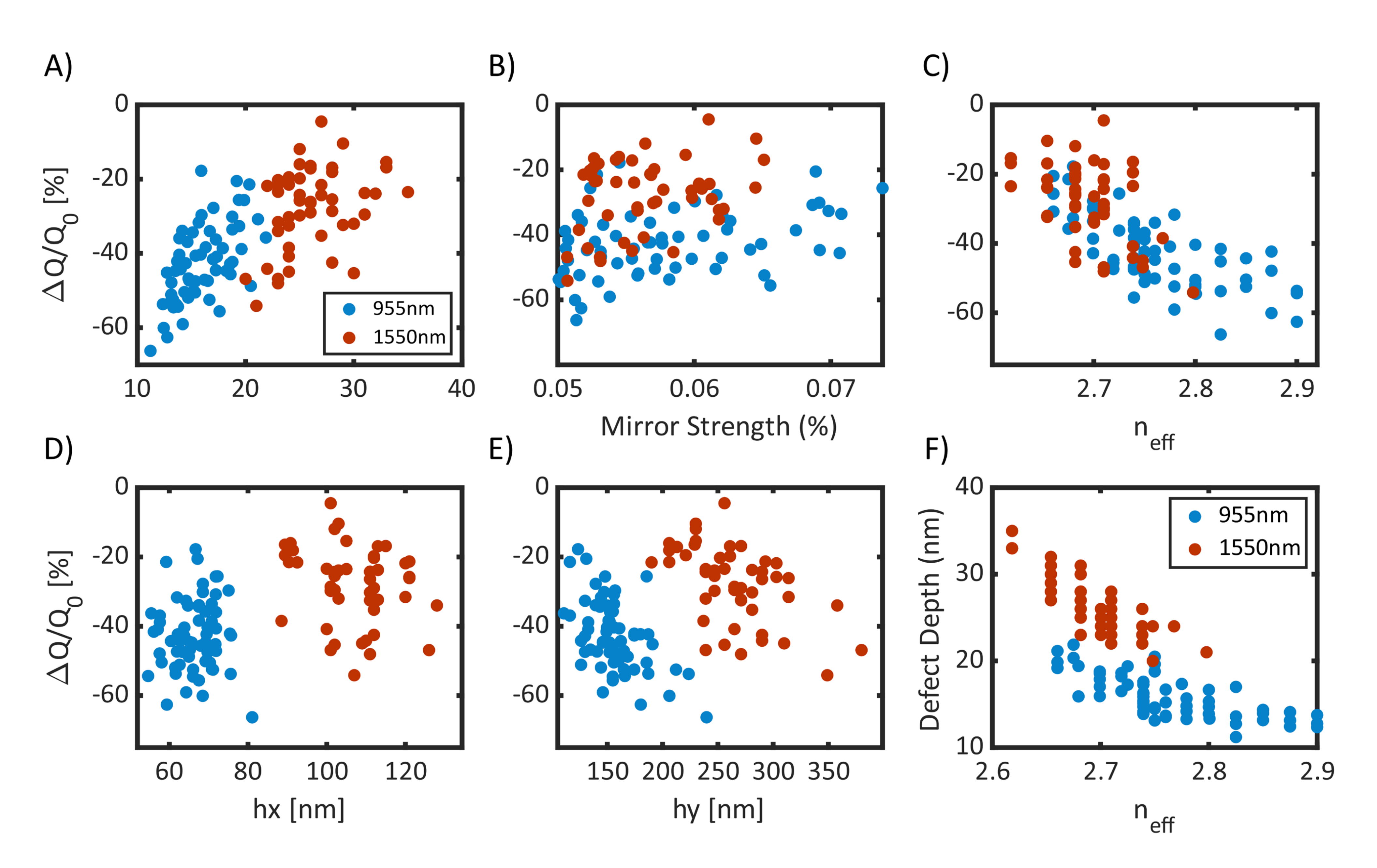}
    \caption{Fabrication error sensitivity as a function of different cavity parameters. Simulated reduction in  quality factor as a function of A) defect depth, B) mirror strength, C) target index, D) hole minor diameter, and E) hole major-diameter. F) Defect depth as a function of target index.
    }
    {
    \phantomsubcaption\label{SI5a}
    \phantomsubcaption\label{SI5b}
    \phantomsubcaption\label{SI5c}
    \phantomsubcaption\label{SI5d}
    \phantomsubcaption\label{SI5e}
    \phantomsubcaption\label{SI5f}
    }
    \label{fig:SI_5}
\end{figure}

To understand the design-to-design variations in robustness, we analyze the simulated reduction in quality factor as a function of different cavity parameters for the 955~nm and 1550~nm cavities as shown in figure \ref{fig:SI_5}. We observe that the robustness to the simulated errors primarily correlates with the defect depth of the cavities as shown in figure \ref{SI5a}. This poses a trade-off however as the defect depth of a cavity appears to anti-correlate with the effective index as shown in figure \ref{SI5f}, which in turn leads to an increase in sensitivity with higher effective indices as shown in figure \ref{SI5c}. Achieving designs with high effective indices and large defect depths can allow for simultaneously high nominal Q and low sensitivity to error. 

\section{Additional Fabrication Details}
\label{S.6}

To fabricate the hybrid GaAs-on-diamond photonic crystal cavities studied in this work, we utilize a stamp transfer process in which the devices are fabricated off chip before transferring onto the diamond as shown in figure 4 of the main text. We begin with an epitaxial GaAs-Al$_{0.8}$Ga$_{0.2}$As-GaAs wafer stack where the top GaAs layer is grown to our target device layer thickness, and the $2um$ thick Al$_{0.8}$Ga$_{0.2}$As layer acts as a sacrificial layer. For the designs fabricated in this work, we use a device thickness of $w_z$ = 220nm. Following solvent cleaning of the chip, we deposit a 5~nm SiO$_2$ adhesion layer at 250$^{\circ}$C using an ICP-CVD (Oxford PlasmaPro 100). We then spin a layer of hydrogen silsesquioxane (HSQ) e-beam resist (Dischem H-SiQ 4\%) at 6000~rpm to yield an 80~nm thick resist layer. The HSQ is stored in liquid nitrogen, and is allowed to thaw for 45 minutes before spinning. After spinning, the sample is dehydration baked at 80$^{\circ}$C for 4 minutes prior to exposure. The patterns are then written into the resist layer at 100KeV (Raith EBPG5150+). The nanobeams are written using a 200~pA, 200~um aperture beam (2~nm spot size) at a dose of 5700~uC/cm$^2$, while the frames and gratings are written with a 20~nA, 300~um aperture beam (13~nm spot size) at a dose of 6500~uC/cm$^2$. A larger beam is used for the frames to reduce write time, while the smaller beam used for the cavities allows for higher resolution. The salty-TMAH developer is prepared by mixing 0.2g of NaCl in 20mL TMAH (25\% in water). The mixture is sonicated for 10 minutes to fully dissolve, then heated to 30$^{\circ}$C on a hotplate. The device chip is baked on a hotplate at 200$^{\circ}$C for 2 minutes, then transferred into the salty TMAH mixture for four minutes to develop with gentle agitation. The chip is placed into successive water beakers followed by IPA, then blow-dried with N$_2$ before a final post-development bake at 200$^{\circ}$C for four minutes. 

\subsection{Improving HSQ-Contrast using Salty-TMAH}
\label{S.6.1}

\begin{figure}[h!tbp]
    \centering
    \includegraphics[width=1\linewidth]{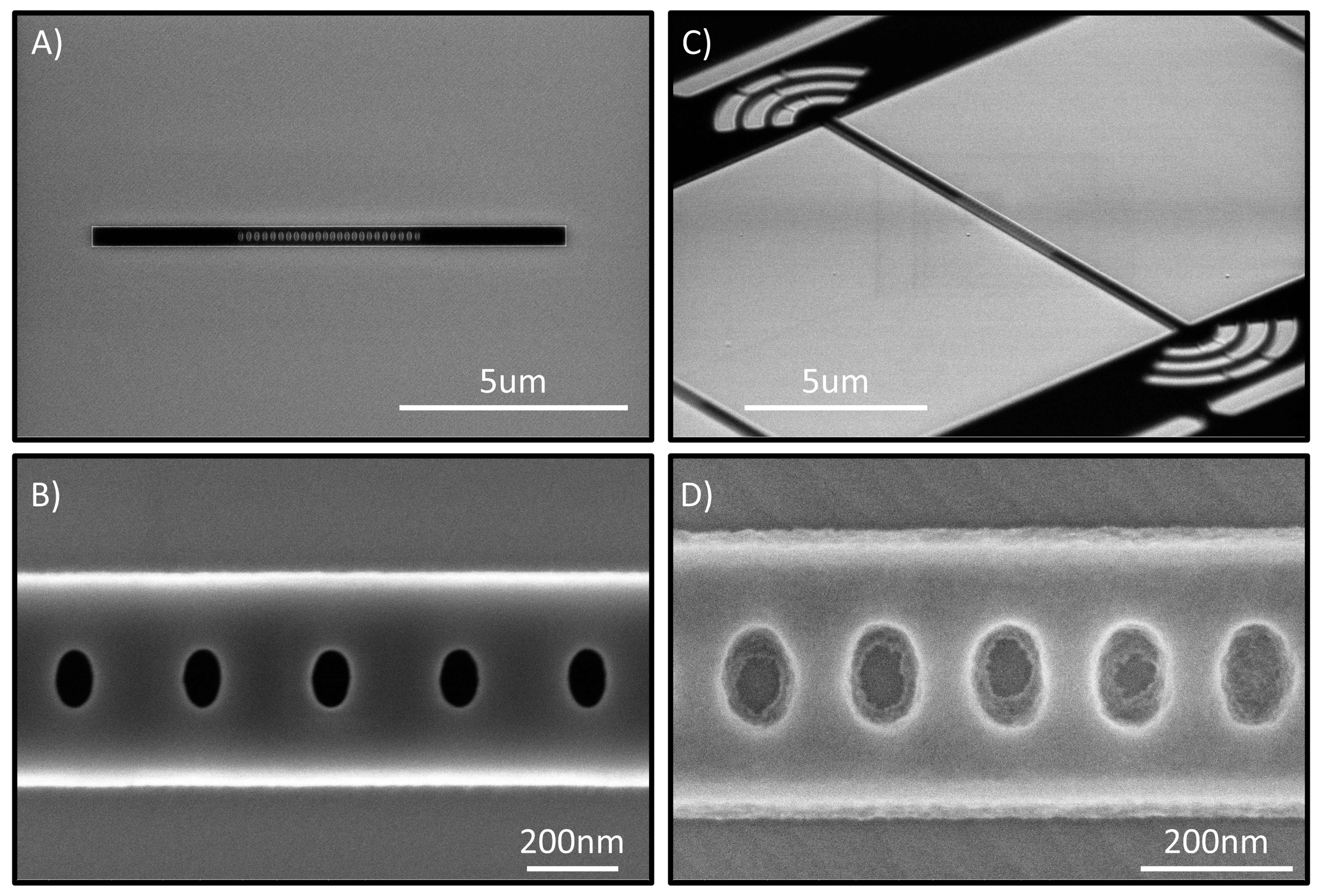}
    \caption{Impact of frame-writing on cavity hole overdosing for devices developed using standard TMAH.
    A) SEM image of an isolated PhC waveguide pattern written in HSQ on GaAs.
    B) Zoom-in image of the isolated PhC waveguide, demonstrating holes as small as $(h_x, h_y) = (70nm, 130nm)$.
    C) A nanobeam written with the surrounding gratings and frame-structures necessary for suspending devices prior to stamp transfer. The PhC waveguide is 10um long.
    D) Zoom-in image of the PhC waveguide, illustrating overdosing within the PhC holes due to electron scatter from the frame write.
    }
        {
    \phantomsubcaption\label{SI6_1a}
    \phantomsubcaption\label{SI6_1b}
    \phantomsubcaption\label{SI6_1c}
    \phantomsubcaption\label{SI6_1d}
    }
    \label{fig:SI_6_1}
\end{figure}

The addition of salt to the TMAH developer is critical for improving the resist contrast and mitigating HSQ development within the cavity holes. In figure \ref{fig:SI_6_1}, we illustrate the impact of secondary electron scatter on the writing of a nanobeam. As we use a stamp transfer fabrication approach, the nanobeams are contained within a frame structure which is used for suspending and later transferring the devices. Despite the large distance between the nanobeam and the frame, and the use of proximity error correction, secondary electron scatter from the frame write results in significant overdosing within the nanobeam holes, setting a lower bound on the achievable hole diameters at $h_x$ = 85nm. By removing the frame, holes with minor diameters as small as $h_x$ = 70nm can be written. As such, elongating the nanobeams within the frames can mitigate the impact of the frame write, but this comes at the expense of reduced pattern density and device stability during the stamp transfer process. As such, we seek to improve the achievable resolution in the presence of secondary scatter from the surrounding features.

\begin{figure}[h!tbp]
    \centering
    \includegraphics[width=0.9\linewidth]{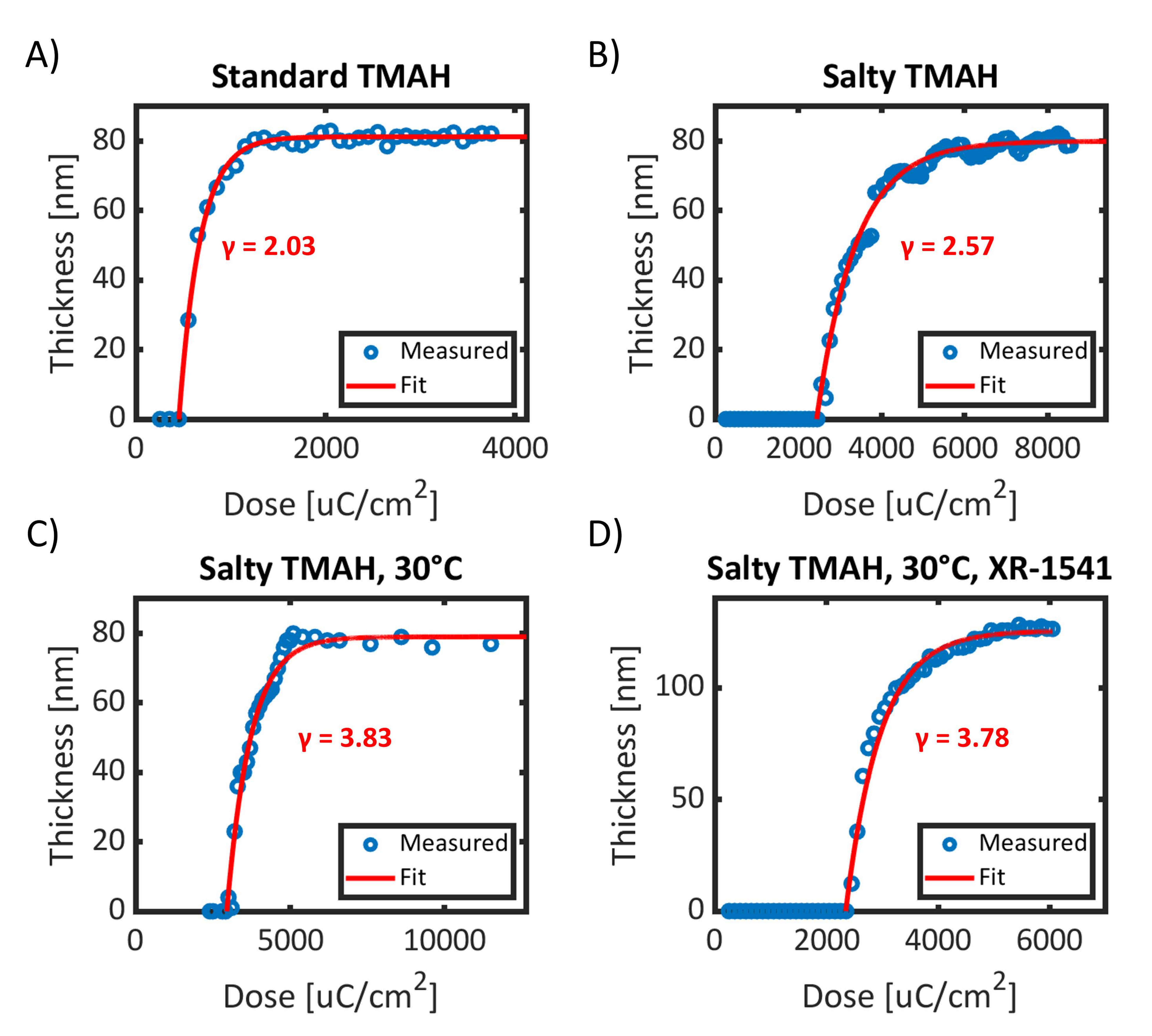}
    \caption{Measured contrast curves for Dischem H-SiQ 4\% developed with A) standard TMAH, B) room-temperature salty-TMAH, and C) heated salty-TMAH. D) Contrast curve for Dow-Corning XR-1541-006 HSQ developed in heated salty-TMAH. The measured data are fit to the model described in equation \ref{EQN_SI6_Resist_Thickness_vs_Dose} and the contrast is extracted according to equation \ref{EQN_Si6_Resist_Contrast}. 
    }
        {
    \phantomsubcaption\label{SI6_2a}
    }
    \label{fig:SI_6_2}
\end{figure}

Previous work has shown that the addition of salt to TMAH can significantly improve the resist contrast by modifying the dissolution rate \cite{yang_using_2007, kim_understanding_2009, yan_effects_2010}. To verify these results, we analyze the resist contrast as a function of development condition as shown in figure \ref{fig:SI_6_2}. To construct the thickness-dose curves, we pattern 20~um x 20~um squares of HSQ at different doses, then measure the resulting thickness using a profilometer (KLA-Tencor P15). The measured data can be fit according to the following equation \cite{mack_data_1999}

\begin{equation}
t(E) = t_0 - \Delta t_{max} exp(-E/E_n^*)
\label{EQN_SI6_Resist_Thickness_vs_Dose}
\end{equation}
where $t(E)$ is the resist thickness after development at a given dose $E$, $t_0$ is the saturated resist thickness, $\Delta t_{max}$ is the maximum resist thickness post development, and $E_n^*$ is a resist sensitivity term. By fitting the measured resist thickness data to this model, we can extract the resist contrast, $\gamma$, as 

\begin{equation}
\gamma \equiv -ln(t_0/\Delta t_{max})
\label{EQN_Si6_Resist_Contrast}
\end{equation}

We observe that the addition of salt increases the resist contrast, and raising the temperature of development further increases the contrast. Additionally, we observe similar results using an alternative HSQ source, Dow-Corning XR-1541-006. 

\subsection{ARDE-Optimized Etching}
\label{S.6.2}
To transfer the patterns into the GaAs device layer, we develop an etch recipe optimized for mitigating the effects of aspect ratio dependent etching (ARDE). As discussed in the main text, ARDE is a phenomenon in which the etch rate within a structure depends on its aspect-ratio, defined as the depth of the structure divided by the width. To develop an ARDE-optimized recipe, we begin with a high-flow, rapid etch as described in table \ref{Table_S9_EtchRecipe_Rapid}. The rapid etch rate achieves high selectivity and vertical sidewalls outside of the PhC holes but the etch rate is severely reduced inside the hole and suffers a high degree of isotropic etching as shown in figure 5 of the main text. To mitigate these effects, we focus on minimizing the etch-rate by reducing ICP power, pressure, and gas flow rates to the lower limits supported by the ICP-RIE tool (Plasma-Therm Takachi). The ARDE-optimized etch is described in table \ref{Table_S9_EtchRecipe}.

\begin{table}[h!tbp]
\centering
\begin{adjustbox}{max width=\linewidth}
\begin{tabular}{|l|l|l|l|}
\hline
 & Step 1 - Gas Stabilization & Step 2 - Etch \\ \hline
Time (s) & 60 & 30 \\ \hline
RF Power (W) & 0 & 50 \\ \hline
ICP Power (W) & 0 & 500 \\ \hline
Chamber Pressure (mTorr) & 5 & 5 \\ \hline
Cl$_2$ Flow Rate (sccm) & 10 & 10 \\ \hline
Ar Flow Rate & 10 & 10 \\ \hline
BCl$_3$ Flow Rate & 10 & 10 \\ \hline
N$_2$ Flow Rate & 10 & 10 \\ \hline
\end{tabular}
\end{adjustbox}
\caption{Rapid etch recipe. The etch parameters are chosen as a baseline which achieves vertical sidewalls outside of the PhC holes with high selectivity. The etch is performed at a chuck temperature of 20$^{\circ}$C.}
\label{Table_S9_EtchRecipe_Rapid}
\end{table}

\begin{figure}[h!tbp]
    \centering
    \includegraphics[width=1\linewidth]{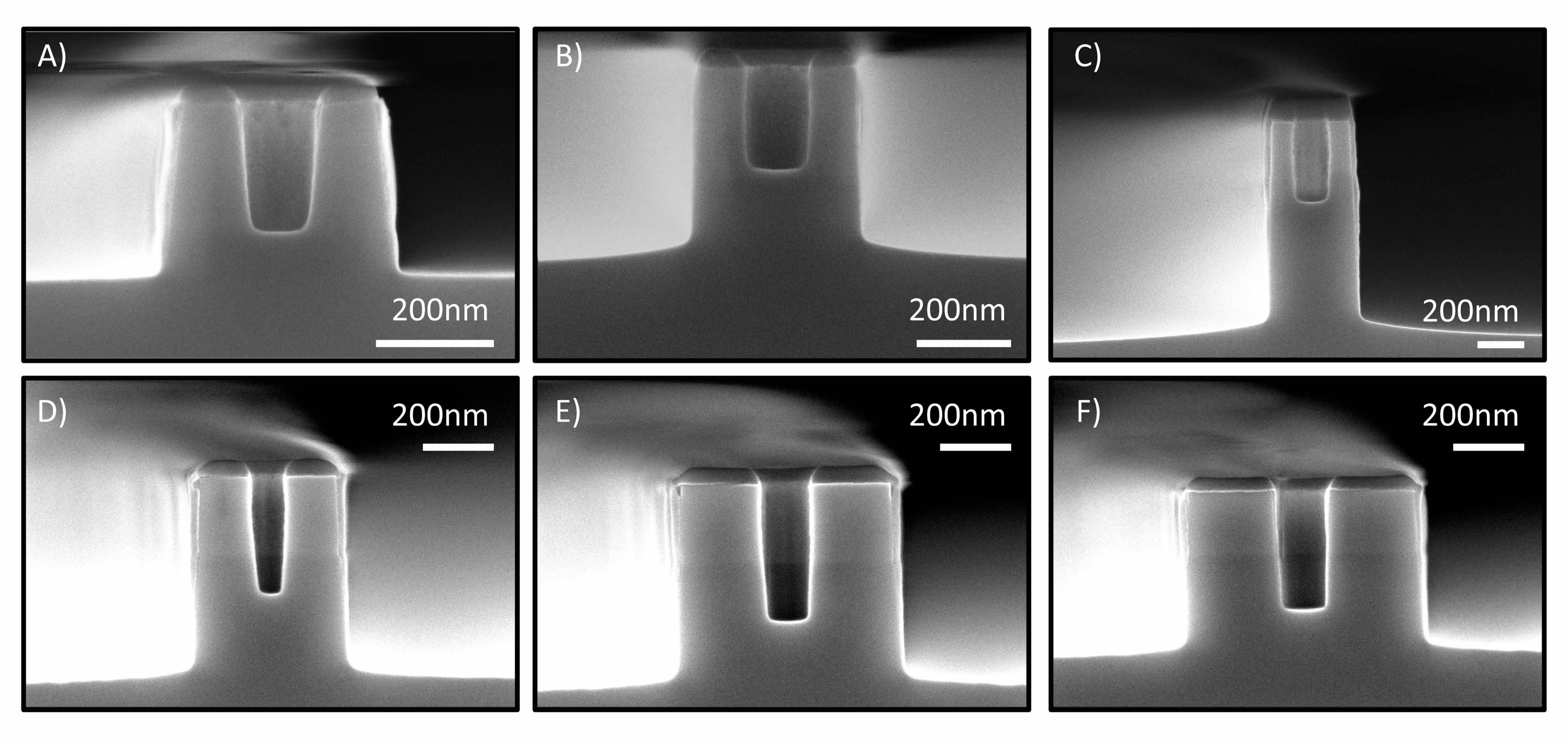}
    \caption{ARDE etch recipe optimization
    A) Cross-sectional SEM of a PhC hole with $h_x$ = 70nm and $h_y$ = 130nm using the etch parameters in table \ref{Table_S9_EtchRecipe} but with 1sccm of Cl$_2$. The etch is overly passivated, resulting in angled sidewalls.
    B) By increasing the Cl$_2$ concentration to 1.75 sccm, we achieve vertical sidewalls within the hole.
    C) Further increasing Cl$_2$ to 3sccm results in higher ARDE and increased isotropic etching.
    D) Cross-sectional SEM of circular holes with diameters of 100nm, E) 150nm, F) 200nm etched with the ARDE-optimized recipe. Reducing the hole-diameter exacerbates the ARDE and increases the degree of isotropic etching within the holes.
    }
        {
    \phantomsubcaption\label{SI6_4a}
    \phantomsubcaption\label{SI6_4b}
    \phantomsubcaption\label{SI6_4c}
    \phantomsubcaption\label{SI6_4d}
    \phantomsubcaption\label{SI6_4e}
    \phantomsubcaption\label{SI6_4f}
    }
    \label{fig:SI_6_4}
\end{figure}

The presence of Cl$_2$ acts to chemically etch the GaAs with a high degree of isotropy, while N$_2$ acts as a passivating agent to prevent isotropic etching. As the minimum flow rate of our tool is 1~sccm for Cl$_2$ and N$_2$, we use this as a starting point for the etch. As shown in figure \ref{SI6_4a}, the resulting etch profile demonstrates a significant reduction in ARDE as compared to the fast etch, but is overly passivated as indicated by the outwardly sloped sidewalls within the hole. To achieve vertical sidewalls, we seek to reduce the passivation. As we are at the lower limit, we can not reduce N$_2$ and thus must increase Cl$_2$ to adjust the balance between etching and passivation. Increasing the Cl$_2$ flow rate to 1.75~sccm results in vertical sidewalls within and outside the hole, but the increased etch rate results in slightly higher ARDE as shown in figure \ref{SI6_4b}. Further increasing the amount of Cl$_2$ leads to a significantly faster etch-rate, more ARDE, and increased isotropy as shown in figure \ref{SI6_4c}. By fine-tuning the relative concentrations, we can achieve vertical sidewalls within the PhC holes for a given design. Importantly, the degree of ARDE varies according to the size of the hole as shown in figures \ref{SI6_4d} - \ref{SI6_4f}, and so the etch-recipe must be optimized according to the smallest hole being fabricated.

\begin{table}[h!tbp]
\centering
\begin{adjustbox}{max width=\linewidth}
\begin{tabular}{|l|l|l|l|}
\hline
& Step 1 - Gas Stabilization & Step 2 - Strike & Step 3 - Etch \\ \hline
Time (s) & 60 & 3 & 330 \\ \hline
RF Power (W) & 0 & 25 & 25 \\ \hline
ICP Power (W) & 0 & 50 & 50 \\ \hline
Chamber Pressure (mTorr) & 2 & 5 & 2 \\ \hline
Cl$_2$ Flow Rate (sccm) & 1.75 & 1.75 & 1.75 \\ \hline
Ar Flow Rate & 2 & 2 & 2 \\ \hline
N$_2$ Flow Rate & 1 & 1 & 1 \\ \hline
\end{tabular}
\end{adjustbox}
\caption{ARDE-optimized etch recipe. The ICP and RF powers are reduced to the lowest values able to support a stable plasma at a pressure of 2 mTorr. The N$_2$ is set to the minimum value according to the tool mass-flow controllers (MFC), while the Cl$_2$ flow is adjusted to achieve vertical sidewalls within the PhC holes. Argon is set to the minimum supported by the MFC and is critical to maintaining plasma stability. The etch is performed at a chuck temperature of 20$^{\circ}$C.}
\label{Table_S9_EtchRecipe}
\end{table}

\subsection{Undercutting and stamp transfer}
\label{S.6.3}
\begin{figure}[h!tbp]
    \centering
    \includegraphics[width=1\linewidth]{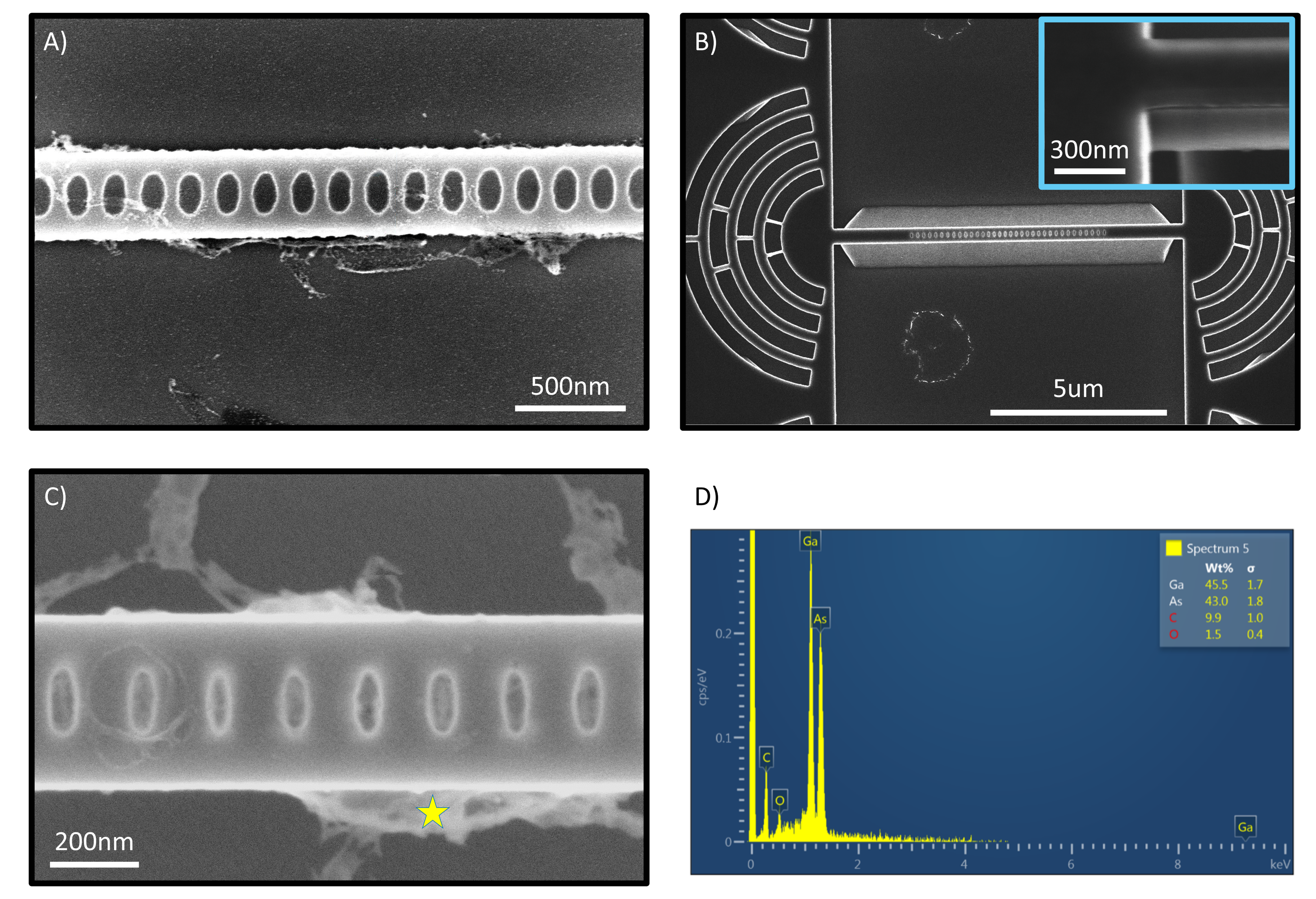}
    \caption{
    A) Residue observed about the PhC after undercutting in HF. Sample was undercut in 1:4 (HF : DI water) for 65s.
    B) Device undercut with concentrated HCl, with no observable residue about the devices.
    C) Device undercut in 1:2 (HCl : DI water) for 8 minutes, followed by a dilute 1:15 (HF : DI water) HSQ-strip for 60s. The HCl undercut is performed in an ice-bath to improve selectivity of the etch.
    D) EDS spectra of the residue obvserved after HCl undercut and dilute HF HSQ-strip. The residue is organic in nature, and can be removed with solvent cleaning.
    }
            {
    \phantomsubcaption\label{SI6_5a}
    \phantomsubcaption\label{SI6_5b}
    \phantomsubcaption\label{SI6_5c}
    \phantomsubcaption\label{SI6_5d}
    }
    \label{fig:SI_6_5}
\end{figure}

\begin{figure}[h!tbp]
    \centering
    \includegraphics[width=1\linewidth]{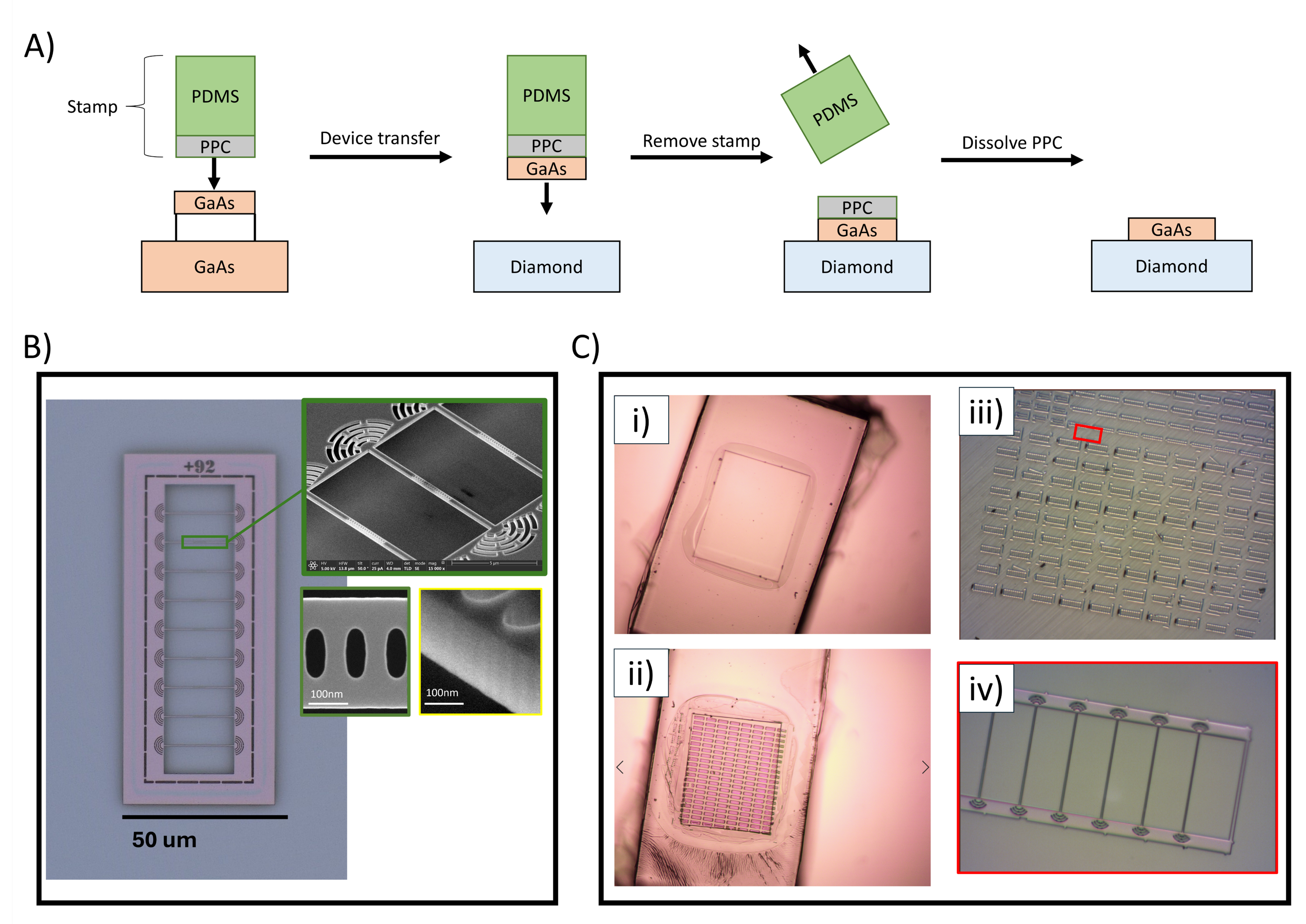}
    \caption{A) Overview of the stamp transfer technique for placing GaAs PhCs onto diamond.
    B) Optical microscope image of a set of devices after undercutting. A given frame supports 7 to 11 identical devices. The shaded pink regions are free-standing, while the thin ring of gray about the outer frame represents the remnant AlGaAs which supports the structure. Insets show typical SEM images of the devices prior to stamp transfer, showing no residue after solvent cleaning.
    C) Stamp transfer process. Inset i) shows an example PDMS stamp with a PPC film over-top. The protudring rectangular region is the mesa which defines the area of device pick-up. ii) Optical image of a set of device patterns after pick-up with the stamp. iii) Devices on diamond after stamping and dissolving of the PPC film. iv) Example image of an in-tact frame of devices on diamond.}
                {
    \phantomsubcaption\label{SI6_6a}
    \phantomsubcaption\label{SI6_6b}
    \phantomsubcaption\label{SI6_6c}
    \phantomsubcaption\label{SI6_6d}
    }
    \label{fig:SI_6_6}
\end{figure}

Following the dry etch, the samples are immediately transferred into IPA to prevent oxidation of the exposed Al$_{0.8}$Ga$_{0.2}$As layer. The patterns are then undercut prior to stamping. While both hydrochloric acid (HCl) and hydrofluoric acid (HF) can be used to etch the sacrificial layer \cite{cheng_epitaxial_2013}, we observe that using HF results in significant residue around the devices as shown in figure \ref{SI6_5a}. In a previous work, the residue was believed to be organic in nature, stemming from polymerization of the e-beam resist during ICP-RIE \cite{kirsanske_electrical_2016}. However, the observed residue is not removed by solvent cleaning, oxygen descum, additional acid soaks, or base soak. The residue can be removed by digital etching techniques \cite{huang_building_2021}, but we observe this to result in unpredictable expansion of the PhC holes and narrowing of the nanobeam. The residue occurs for both ZEP and HSQ e-beam resists, and is not observed after etching prior to undercutting. Alternatively, HCl does not produce any residue when undercutting the sacrificial layer as shown in figure \ref{SI6_5b}. This confirms that the residue is specific to some reaction between the HF and sacrificial layer. When using HCl to undercut the devices, HF must eventually be used to remove the HSQ, which results in a new form of residue as shown in figure \ref{SI6_5c}. However, as the HCl has removed the majority of the AlGaAs, the residue following HSQ stripping is not likely to be related to the HF undercut issue. EDS measurements confirm the residue to be organic, and soaking the chip in solvents can remove the residue, resulting in pristine devices as shown in figure \ref{SI6_6b}. The solvent-cleaning procedure consists of soaking in Microchem Remover PG (Kayaku) at 80$^{\circ}$C for 24 hours, followed by a 3 hour soak in acetone at 50$^{\circ}$C. Following the solvent cleaning, the sample is transferred to IPA then dried using a critical point dryer (Baltec-030).

Finally, the cavities are transferred to diamond using a stamp transfer process as shown in figure \ref{SI6_6a} and described in \cite{dibos_atomic_2018}. Using the stamp transfer technique, we transfer device areas as large as 2mm x 2mm, corresponding to up 4000 individual cavities. Prior to stamp transfer, the diamond surface is cleaned in a 2:1 piranha mixture and dehydration baked at 100$^{\circ}$C for 10 minutes. 

\newpage
\section{Measurement Setup}
\label{S.7}

\begin{figure}[h!tbp]
    \centering
    \includegraphics[width=1\linewidth]{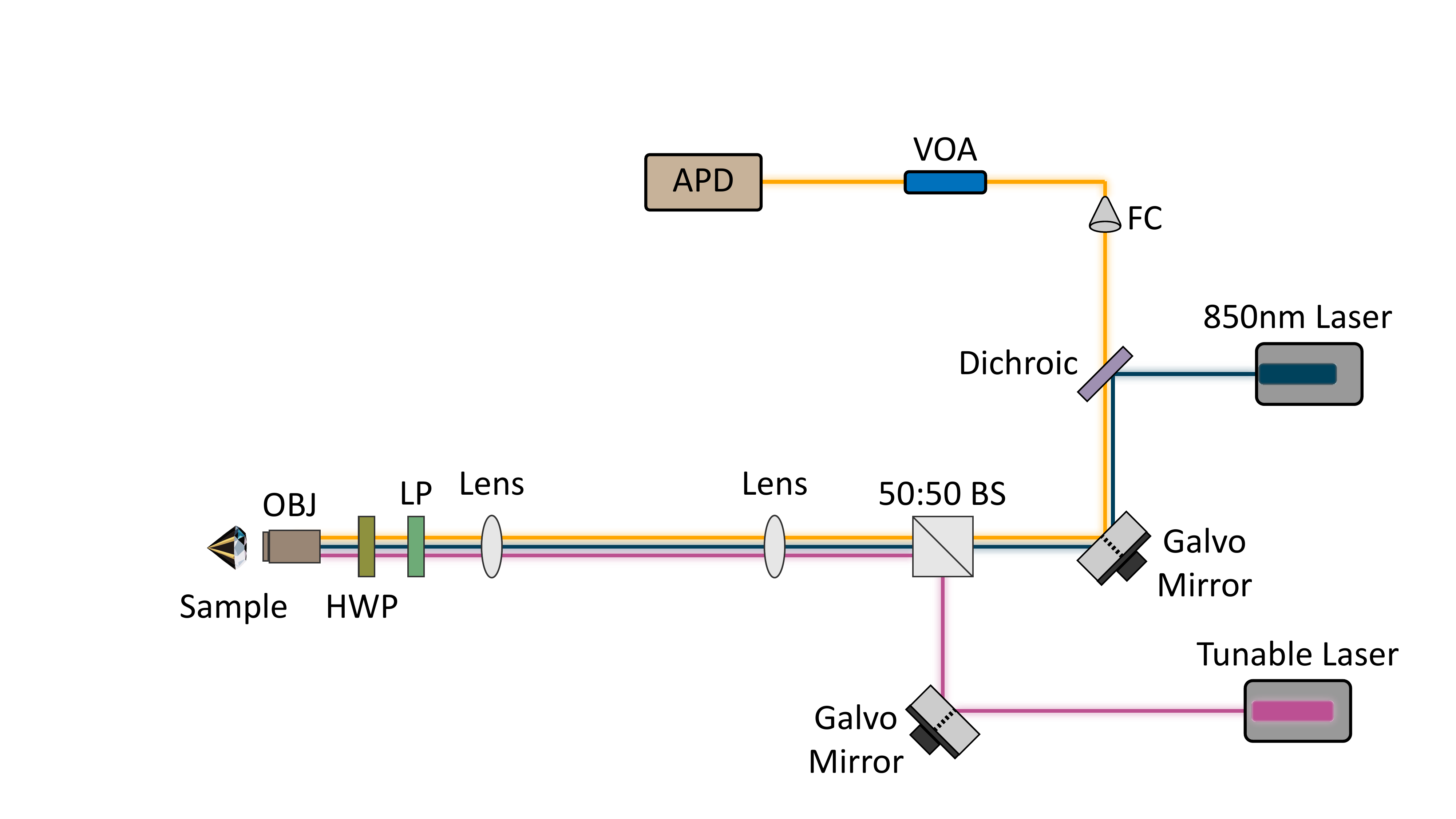}
    \caption{Optical measurement setup used for measuring cavity spectra. Measurements are performed using a free-space 4f confocal measurement setup in conjunction with control software for beam steering and laser control.}
    \label{fig:SI_7}
\end{figure}

The cavities in this work are measured using a confocal microscopy setup as shown in figure \ref{fig:SI_7}. The excitation channel consists of a tunable laser (Toptica CTL-950) steered with a 2-axis scanning galvo mirror (Thorlabs GVS012) through a 4f relay system. The beam passes through a linear polarizer and motorized half-wave plate into a 0.85 NA objective (Olympus LCPLN100XIR). The polarization optics allow for selective measurement of the cavity TE spectrum. To measure the transmitted light through the cavity, the detection channel is independently steered using an additional galvo mirror, and is combined with the excitation through the 4f system using a 50:50 cube beamsplitter. A silicon APD is used in conjunction with a voltage controlled attenuator for detecting counts. To image the devices under the objective, an additional 850~nm diode laser is included in the detection path which allows for confocal imaging across the objective field of view. From this confocal image, the individual devices and gratings can be identified, and control software is used to automatically track the in and out-coupling spots on the gratings, and to sweep the laser wavelength. For measurement of high-Q devices, a scanning Fabry-Perot interferometer (Thorlabs SA30-95) is used to accurately calibrate the resonance linewidths. For measurements in the telecom band, the excitation laser is replaced with a Santec TSL-770, and the silicon APD is replaced with a superconducting nanowire single-photon detector (SNSPD).

\newpage
\section{Additional Cavity Measurements}
\label{S.8}

\begin{figure}[h!tbp]
    \centering
    \includegraphics[width=1\linewidth]{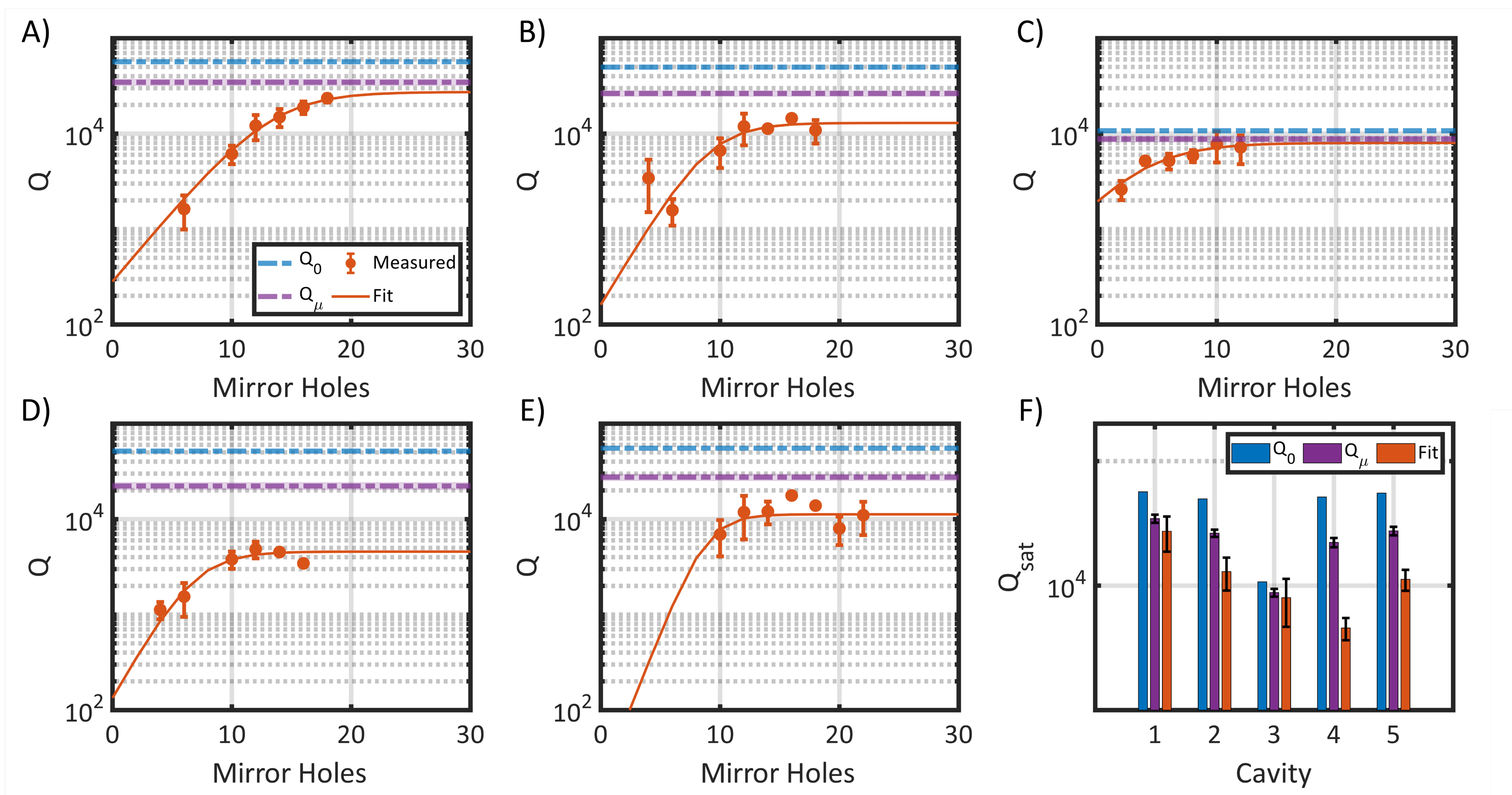}
    \caption{Q-scaling measurements of the five cavities used to verify the fabrication-error simulation model. A) - E) Measured data plotted against the nominal and noisy simulations for five different cavity designs. F) Comparison between the nominal simulated Q-factor, noisy simulated Q-factor, and fitted intrinsic Q.}
    {
    \phantomsubcaption\label{SI8a}
    \phantomsubcaption\label{SI8b}
    \phantomsubcaption\label{SI8c}
    \phantomsubcaption\label{SI8d}
    \phantomsubcaption\label{SI8e}
    \phantomsubcaption\label{SI8f}
    }
    \label{fig:SI_8}
\end{figure}

In figure \ref{fig:SI_8} we show the Q-scaling measurements of the five cavities plotted in figure 6 of the main text. For each cavity, we fit the measured data to find the intrinsic quality factor of the cavity. The fitted intrinsic Q is compared to the nominal and noisy simulations to benchmark the accuracy of the fabrication-error simulations.

\section{Measurement Results of High-Q Designs}
\label{S.9}

Given the success of the noisy-simulation model, we attempt to apply our design rules to cavities with higher nominal Q-factors. We select cavity designs with a range of simulated sensitivity to fabrication errors, but with quality factors in excess of 100,000. In measuring the cavities however, we observe a breakdown of the fabrication-error model as all designs saturate at quality factors of 30,000. The origin of this saturation can arise from either material losses or fabrication-induced losses.

\begin{figure}[h!tbp]
    \centering
    \includegraphics[width=1\linewidth]{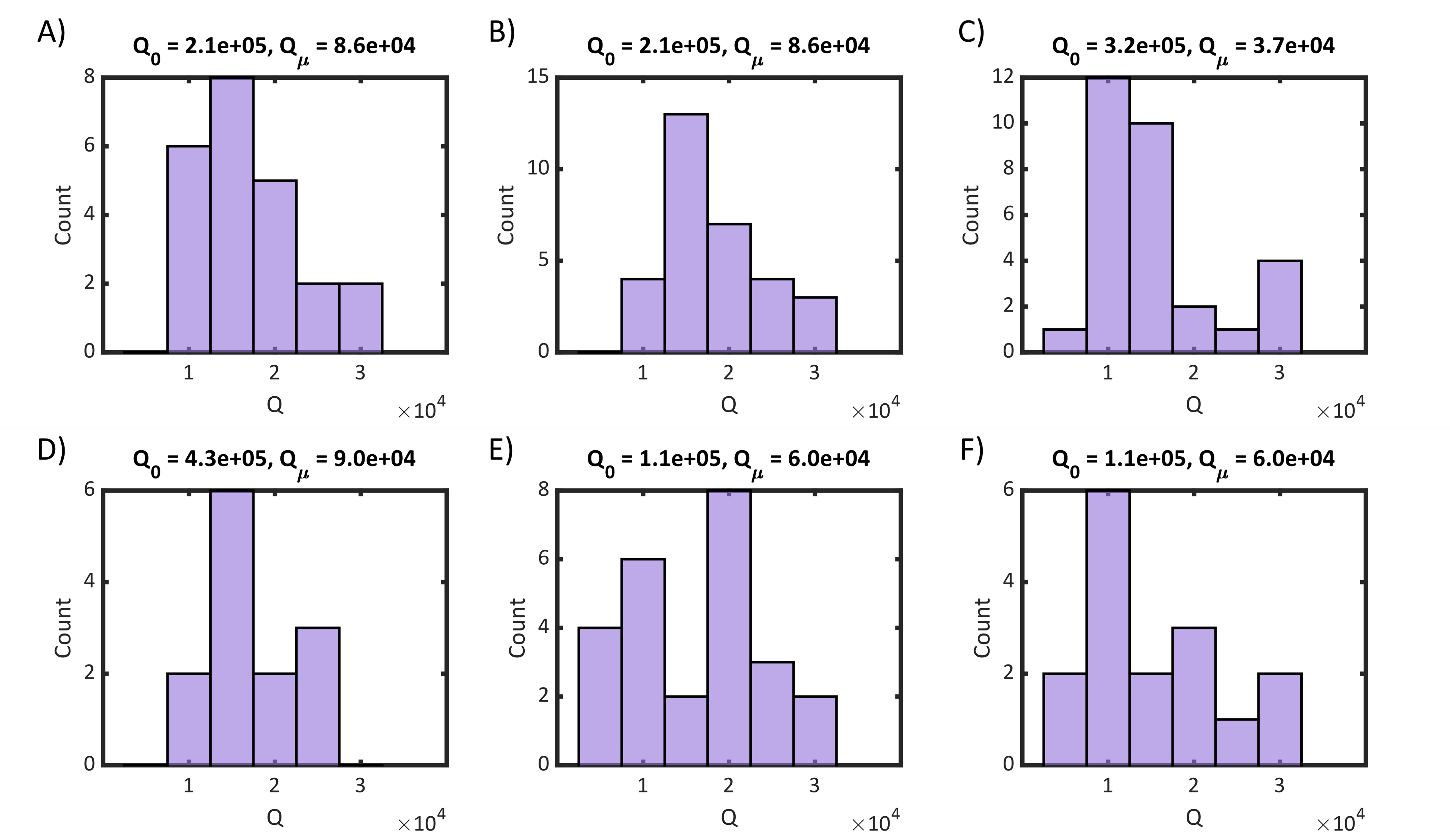}
    \caption{Measurement results of several high-Q designs showing saturation at Q=30,000. A)-F) Measurement results of six different cavity designs with a range of simulated nominal Q and fabrication sensitivity. The Q=30,000 limit across designs exemplifies the breakdown of our fabrication-sensitivity model, indicating that an alternative source of error is limiting device Q.}
    {
    \phantomsubcaption\label{SI9a}
    \phantomsubcaption\label{SI9b}
    \phantomsubcaption\label{SI9c}
    \phantomsubcaption\label{SI9d}
    \phantomsubcaption\label{SI9e}
    \phantomsubcaption\label{SI9f}
    }
    \label{fig:SI_9}
\end{figure}

\section{Measurement Results using MOCVD GaAs Wafers}
\label{S.10}
\begin{figure}[h!tbp]
    \centering
    \includegraphics[width=0.5\linewidth]{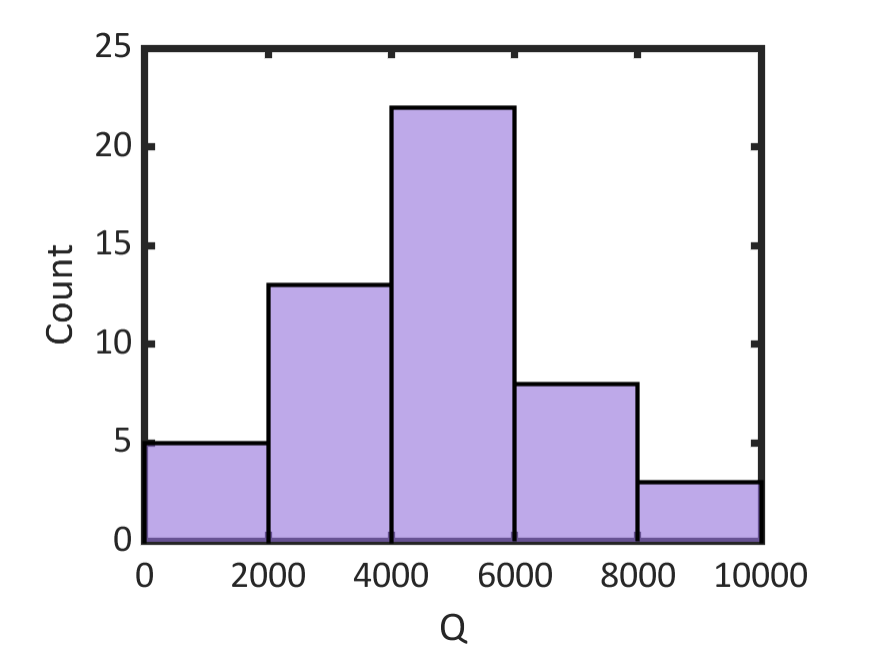}
    \caption{Measurement results of several high-Q designs fabricated using a commercial MOCVD wafer. Plotted are the results for three different cavity designs, all with nominal Q exceeding 100,000. For all fabricated cavities, a maximum quality factor of 10,000 is observed using the MOCVD wafers.}
    \label{fig:SI_10}
\end{figure}

In figure \ref{fig:SI_10} we plot the measurement results for cavities fabricated using a commercial MOCVD GaAs wafer. In contrast to the MBE wafers used in the main text, we observe much lower quality factors. For the MOCVD wafers, the maximum achievable quality factors are below 10,000.

\newpage
.
\newpage
\bibliographystyle{ieeetr} 
\bibliography{supplementary}